\documentclass[twocolumn,showpacs,nofootinbib,preprintnumbers,amsmath,amssymb]{revtex4}
\usepackage{graphicx}
\usepackage{epsfig}
\usepackage{amssymb}
\usepackage{dcolumn}
\usepackage{bm}
\usepackage[hypertex]{hyperref}
\usepackage{psfrag}

\newcommand{\be}{\begin{equation}}
\newcommand{\ee}{\end{equation}}
\newcommand{\bea}{\begin{eqnarray}}
\newcommand{\eea}{\end{eqnarray}}
\newcommand{\la}{\left<}
\newcommand{\ra}{\right>}

\begin{document}
\title{$1-d$ gravity in infinite point distributions}

\author{A. Gabrielli$^{1,2}$, M. Joyce$^{3}$, F. Sicard$^{3}$}

\affiliation{$^1$SMC, INFM/CNR, Physics Department,\\
University ``La Sapienza'' of Rome, 00185-Rome, Italy}
\affiliation{$^2$Istituto dei Sistemi Complessi,\\ 
CNR, Via dei Taurini 19, 00185-Rome, Italy}
\affiliation{$^3$Laboratoire de Physique Nucl\'eaire et Hautes \'Energies,\\ 
Universit\'e Pierre et Marie Curie - Paris 6,
CNRS IN2P3 UMR 7585, Paris,F-75005, France}

\begin{abstract}   
The dynamics of infinite, asymptotically uniform, distributions
of purely self-gravitating particles in one spatial dimension
provides a simple and interesting toy model for the analogous 
three dimensional problem treated in cosmology. In this article
we focus on a limitation of such models as they have been treated 
so far in the literature: the force, as it has been specified, 
is well defined in infinite point distributions only if
there is a centre of symmetry (i.e. the definition requires
explicitly the breaking of statistical translational invariance). 
The problem arises because naive background subtraction (due to 
expansion, or by ``Jeans swindle'' for the static case), applied 
as in three dimensions, leaves an unregulated contribution to the force
due to surface mass fluctuations.  Following a discussion 
by Kiessling of the Jeans swindle 
in three dimensions, we show that the problem may be resolved by 
defining the force in infinite point distributions as
the limit of an exponentially screened pair interaction. We 
show explicitly that this
prescription gives a well defined (finite) force acting on particles 
in a class of perturbed infinite lattices, which 
are the point processes relevant to cosmological $N$-body
simulations. For identical particles the dynamics of the simplest 
toy model (without expansion) is equivalent to that of an infinite 
set of points with inverted harmonic oscillator potentials which bounce 
elastically when they collide. We discuss and compare with previous results
in the literature, and present new results for the specific 
case of this simplest (static) model starting from ``shuffled lattice''
initial conditions. These show qualitative properties of the evolution
(notably its ``self-similarity'') like those in the analogous simulations 
in three dimensions, which in turn resemble those in the expanding universe.
\end{abstract}


\maketitle

\section{Introduction}
The development of clustering in initially quasi-uniform {\it infinite}
distributions of point particles evolving purely under their
Newtonian self-gravity has been the subject of extensive numerical 
study in cosmology over the last several decades (see
e.g. \cite{bagla_review} for a review). This is the case because these
``N-body'' (particle) simulations of the Newtonian limit are believed 
to give a very good approximation to the formation of structure 
formation in current dark matter dominated models of the universe. 
The impressive growth in the size of these simulations has 
led essentially to phenomenological models of the 
associated dynamics. Analytical understanding, which would be 
very useful in trying to extend the numerical results and
also control for their reliability, remains very limited. In 
attempts to progress in this 
direction it is natural to look to simplified toy models which
may provide insight and qualitative understanding. Such models
may also be interesting theoretically in a purely statistical
mechanics setting, and specifically in the 
context of the investigation of out of equilibrium dynamics
of systems with long-range 
interactions (see e.g. \cite{Dauxoisetal, Assisi}).

An obvious toy model for this full $3-d$ problem
is the analogous problem in $1-d$, i.e., the generalization 
to an infinite space (static or expanding) of the 
so-called ``sheet model'', which is formulated for 
finite mass distributions. In this latter model, 
which has been quite extensively 
investigated (see, e.g., \cite{hohl+feix, reidl+miller, severne+luwel,
rouet+feix, tsuchiya+gouda, miller_1dreview} and 
references therein), particles in $1-d$
experience pair forces independent of 
their separation, like those between parallel 
self-gravitating  sheets in $3-d$ of infinite extent. 
Several groups of authors \cite{rouet_etal, yano+gouda, 
tatekawa+maeda, aurell_etal, miller+rouet_2002, aurell+fanelli_2002a,
aurell+fanelli_2002b, miller+rouet_2006, 
valageasOSC_1, valageasOSC_2, miller_etal_2007} have then discussed different variants on
this model to develop the analogy with the $3-d$
infinite space problem. Just as for the 
finite sheet model, these models have the particular interest of 
admitting exact solutions between sheet crossings, which means 
that they can be easily evolved numerically to machine precision, and
at modest numerical cost for quite large numbers of particles.

In this article we revisit the basics of these toy models
(in either static or expanding universes), addressing the problem 
of their general formulation for infinite distributions.
Indeed, as we will discuss, previous discussions have required, in 
their implementation, the imposition of symmetry about a point, or 
finite extent of the considered density perturbations\footnote{This is not
true of the treatments in \cite{yano+gouda, tatekawa+maeda}, which
start directly from the fluid limit (rather than from a particle
description). See further discussion below.}.  
Such a restriction on the class of point processes which can
be considered, and notably the requirement that statistical
translational invariance be broken, is not desirable.
Indeed in the context of the cosmological problem, 
this latter property of the distributions 
usually considered as initial conditions for simulations
is very important, because of the ``cosmological principle'' 
which supposes that there are no preferred centres
(see e.g. \cite{peebles, book}). Further the question of 
the extrapolation of the finite version of the model (which 
is what is simulated numerically) to the infinite system 
limit has, as we will discuss below, not been carefully examined.
We will show that problems with the definition of the 
force (as used in these previous treatments) arise from a 
subtlety about how the so-called ``Jeans swindle'' is applied in one
dimension. We draw here on the work of Kiessling in \cite{kiessling},
where it has been shown that, in $3-d$, the
usual formulation of the Jeans swindle --- subtraction
of a compensating negative mass background in calculation
of the potential --- may 
be more physically formulated as a prescription for the
calculation of the {\it force} in the infinite volume limit.
It turns out, as we will see, that while in $3-d$ 
it is sufficient to prescribe that the force on a given
particle is obtained by summing symmetrically about it
(e.g. summing in spheres of radius $R$ with centre at
the particle, and then sending $R$ to infinity), in
$1-d$ this limiting procedure needs to be
further specified. More specifically the force turns
out to be defined in $1-d$ for a broader class of point
distributions --- and notably for distributions
without a centre --- when the summation is 
performed by taking the unscreened limit of 
the same sum for a screened version of the interaction,
rather than as the limit of the sum truncated to 
a finite symmetric``top-hat'' interval.

In the next section
we give a more detailed heuristic discussion of the problem
of defining the force in the infinite volume
limit, and then give the prescription we adopt.
We then present in Sect.~\ref{Forces in infinite perturbed lattices}
a rigorous calculation showing that the force is indeed well 
defined for a certain class of infinite perturbed lattices, i.e., 
infinite configurations generated by perturbing particles 
off a perfect lattice. To do this we treat these infinite 
point distributions as stochastic point processes and study
the probability distribution of the force on 
particles\cite{chandra43, book}.
For finite variance displacements off the lattice,
either correlated or uncorrelated, which do not cause 
particles to cross one another, the result turns
out to be extremely simple: the force on any particle 
is simply proportional to its displacement from its lattice
site.  In the following section we turn to consider the 
definition of dynamical models using this force. While
the most evident model is the simple one obtained by
using the conservative Newtonian dynamics under
the derived forces, there is also a simple variant   
with an additional damping term which is the natural
toy model for the $3-d$ cosmological problem (with
an expanding background).  Given that 
particle crossings are, up to interchanges of particle labels, 
equivalent, in $1-d$, to an elastic collision
(with exchange of velocities) the evolution in these
toy models, starting from infinite perturbed lattices
in the class in which we have shown the force to
be defined, is in fact equivalent {\it at all times} 
simply to an infinite set of particles with 
inverted harmonic oscillator potentials centred
at their original lattice sites, and which collide
elastically when they meet. In the last subsection of
this section,  we then give a detailed discussion of 
the relation of these two toy models to those which
have been discussed previously in the literature,
explaining that our formulation provides essentially
both a simplification, and a generalization, of most
of these previous treatments.
In Sect.~\ref{Case study}
we present briefly results of numerical simulations for 
the simplest toy model (without expansion), and uncorrelated 
initial
displacements to the lattice (the ``shuffled lattice''). 
We show that in this case the evolution
of the clustering in time is qualitatively very similar
to that which has been observed in the analogous $3-d$ 
system. Notably these static space simulations
share the features of ``hierachichal'' structure formation
and ``self-similarity'' which are well documented in 
full $3-d$ simulations. In the final
section we summarize our findings and conclusions, and 
discuss some directions we envisage for further work.

\section{From finite to infinite systems}
\label{From finite to infinite systems}


\subsection{Definitions}

By gravity in one dimension we mean the pair
interaction corresponding to an attractive force independent
of separation, i.e., the force $f(x)$ on a particle at
coordinate position $x$ exerted by a particle at the 
origin is given by 
\begin{equation}
\label{Familiar_pair_force}
 f(x) = -g\frac{x}{\vert x \vert} = -g~\textrm{sgn}(x)~,
\end{equation}
where $g$ is the coupling. Equivalently it is the pair
interaction given by the pair potential $\phi(x) = g|x|$ 
which satisfies the $1-d$ Poisson equation
for a point source, $\frac{d^{2}\phi}{dx^{2}}=2g \delta_{D} (x)$
(where $\delta_D$ is the Dirac delta function). Comparing with 
the $3-d$ Poisson equation
shows the equivalence with the case of an infinitely 
thin plane of infinite extent and surface mass density 
$\Sigma=g/2\pi G$, which explains the widely used
name ``sheet model''. We will work in the one
dimensional language, referring to ``particles''. For
convenience we will set the mass of these particles,
which will always be equal here, to unity. 

\subsection{Finite system}

Let us consider first the case of a finite system, consisting
of a finite number $N$ of particles (with either open 
boundary conditions, or contained in a finite box). 
Denoting by $x_{i}$ the coordinate position 
of the $i^{th}$ particle along the real axis, the force field 
$F(x)$ (i.e. the force on a test particle) at the point $x$ 
is 
\begin{equation}
\label{Force1D-simple}
  F(x) = g \sum_{i}\textrm{sgn}(x_{i}-x) 
  = g \int dy~n(y)~\textrm{sgn}(y-x)~,
\end{equation}
where $n(y) = \sum_{i}\delta_{D}(y-x_{i})$ is the microscopic number  
density and the integral is over the real line\footnote{We use the
standard convention that $\textrm{sgn}(0)=0$, which implies
this same formula is valid for the force on a particle of 
the distribution (rather than a test particle) at $x$.}.  
Equivalently it may be written as 
%
%
\begin{equation}
  F(x) = g \Big[ N_{>}(x)- N_{<}(x) \Big].
\label{force-finite-system}
\end{equation}
where $N_{>}(x)$ ($N_{<}(x)$) is the number of particles to
the right (left) of $x$.
The dynamics of this model, from various initial conditions and
over different times scales, has been extensively explored in the
literature (see references given above). 

\subsection{Infinite system limit}

Let us consider now the infinite system limit, i.e., an
infinite uniform distribution of points\footnote{By ``uniform'' 
we mean that the point process 
has a well defined {\it positive} and spatially constant mean 
density, i.e., it becomes homogeneous at sufficiently large scales.}  
on the real line  with some mean 
density $n_0$ (e.g. a Poisson process). 
It is evident that the forces acting on particles are not 
well defined in this limit, as the difference between
the number of particles on the right and left of a given
particle depends on how the limit is taken. Formally we can 
write the force field of Eq.~(\ref{Force1D-simple}) as
\begin{equation}
\label{Force_Reg}
 F(x) = g n_0 \int dy ~\textrm{sgn}(y-x) + g \int dy~\delta n(y)~\textrm{sgn}(y-x)~,
\end{equation}
where $\delta n(y) = n(y)-n_{0} = \sum_{i}\delta_{D}(y-x_{i})- n_{0}$ represents the
number density fluctuation. While the second term would, naively, 
be expected to converge if the fluctuations $\delta n(y)$ can 
decay sufficiently rapidly, 
the first term, due to the mean density, is explicitly badly defined (as the
integral is only semi-convergent). Precisely the same problem 
arises for gravity in infinite $3-d$ distributions. The solution,
known as the ``Jeans swindle'', is the subtraction of the
contribution due to the mean density. 
As discussed by Kiessling in \cite{kiessling},  rather than 
a ``swindle'', this is, in $3-d$, in fact a mathematically  
well-defined regularisation of the physical 
problem, corresponding simply to the prescription that 
the force at a point be summed so that it vanishes in the limit 
of exact uniformity. 
The simplest form of 
such a prescription in $3-d$ is that the force on a particle be 
calculated by summing symmetrically about the particle (e.g. by summing 
about the considered point in spheres of radius $R$, and then 
sending $R \rightarrow \infty$). This formulation needs no
explicit use of a ``background subtraction'', since the 
term due to the mean density does not contribute when
the sum is performed symmetrically.

Applying the same reasoning to the $1-d$ case would
lead to the prescription 
\begin{equation}
\label{Force_Reg_secondterm}
F(x) = g \int dy~\delta n(y)~\textrm{sgn}(y-x)~.
\end{equation} 
The question is whether this expression for the 
gravitational force is now well defined, and if it
is, in what class of infinite point distributions.
As we will detail in the next section of the paper, this 
question may be given a precise answer, as in $3 - d$,  by 
considering the probability density function of the force in 
such distributions, described as stochastic point
processes in infinite space.  In the rest of this section
we will simply explain the problems which arise when
the infinite system limit of expression 
Eq.~(\ref{Force_Reg_secondterm}) is taken using
a simple top-hat prescription. This discussion motivates
the use of a smooth version of this prescription, which
we then show rigorously in the subsequent section 
to give a well defined force for a broad class of
infinite perturbed lattices.

For Eq.~(\ref{Force_Reg_secondterm}) to be well defined in an
infinite point distribution it must give the same answer
no matter how it is calculated. Two evident top-hat prescriptions
for its calculation are the following. On the one hand it 
may be written as
\begin{equation}
\label{Force_Reg_Inf}
 F(x) = g \lim_{L \rightarrow \infty} \int_{x-L}^{x+L} dy~ n(y)~\textrm{sgn}(y-x)~,
\end{equation}
or, equivalently, 
\begin{equation}
F(x) = g \lim_{L \rightarrow \infty} \Big[ N (x,x+L)- N (x-L,x)
 \Big],
\label{force-top-hat-diff}
\end{equation}
where $N (x,y)$ is the number of points between $x$ and $y$,
i.e., the force is proportional to the difference in the 
number of points on the right and left of $x$ inside 
a {\it symmetric interval centred on} $x$, when the size of
the interval is taken to infinity. On the other hand, we can
write
\begin{equation}
\label{Force_symmTH}
 F(x) = g \lim_{L \rightarrow \infty} \int_{-L}^{+L} dy~ \delta n(y)~\textrm{sgn}(y-x)~,
\end{equation}
or, equivalently, 
\begin{equation}
F(x) = g \lim_{L \rightarrow \infty} \Big[ N (x,L)- N(-L,x) 
 \Big] + 2g n_0 x,
\label{force-symmTH-diff}
\end{equation}
i.e., we integrate the mass density fluctuations in a top-hat centred
{\it on some arbitrarily chosen origin}. 
\begin{figure}[!t]
 \begin{center}
 \includegraphics[width=7cm]{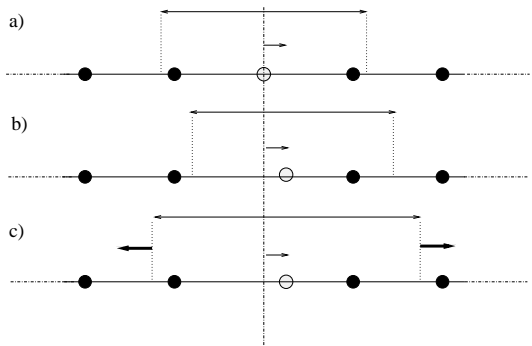}
\end{center}
\caption{Calculation of the force using a top-hat regularisation centred
on the point considered, i.e., as defined in Eq.~(\ref{force-top-hat-diff}).
In an unperturbed lattice (case a) the force
on points of the lattice vanishes. However, as shown in b) and c), 
when a single point is displaced off lattice, the force becomes
badly defined, oscillating between $g$ and zero as the size of 
top-hat goes to infinity.}
\label{SL_example}
\end{figure}

That these expressions are both badly defined in an infinite
Poisson distribution is easy to see: 
in this case the fluctuation in mass on the right of 
any point is uncorrelated with that on the left, 
giving a typical force proportional to the square root 
of the mass in a randomly placed window of size $L$, which 
grows in proportion to $\sqrt{L}$ (and thus diverges). 
Calculating the force with Eq.~(\ref{force-top-hat-diff})
one  of us (AG) has shown in \cite{andrea_1dforces} that
it is in fact not well defined either in a class of 
more uniform distributions of points, randomly 
perturbed lattices\footnote{The 
force is, however, shown to be well defined in this class
of point distributions using the analogous definition for
any power law interaction in which the pair force decays 
with separation. See \cite{andrea_1dforces}
for details.}. Why this is so can be understood easily
by considering, as illustrated in  Fig.~\ref{SL_example}, the 
calculation of the force using Eq.~(\ref{force-top-hat-diff}) 
in such configurations.  While on the unperturbed
lattice (case a) the force on all points of the lattice
is well-defined (and vanishing, as it should be), 
this is no longer true when a particle is displaced:
the force on the displaced particle now oscillates
deterministically (between $g$ in case b, and zero 
in case c) and does not converge 
as $L \rightarrow \infty$. 

For the same case, of a single particle displaced off an infinite
perfect lattice, the prescription Eq.~(\ref{force-symmTH-diff})
for the force does, however, give a well-defined result if one chooses 
as origin a point of the {\it unperturbed} lattice: since the 
first (``particle'') term is unchanged by the displacement of 
the particle, the only non-vanishing contribution comes from 
the second (``background'') term, giving a finite force 
\be F(u)=2gn_0u\;,
\label{force-single-particle}
\ee
where $u$ is the displacement of the particle from its lattice site 
(i.e. the centre of symmetry) and we assume $u$ is smaller than the 
lattice spacing.
If we consider now, however, applying random displacements of 
small amplitude (compared to the interparticle spacing) to the other
particles of the lattice, the problem of the first prescription
Eq.~(\ref{force-top-hat-diff}) reappears: at any given $L$
the first term in Eq.~(\ref{force-symmTH-diff}) picks up
a stochastic fluctuation which varies discretely between 
$\pm g$ and zero, and does not converge as $L \rightarrow \infty$. 
This will evidently be the case for any such configuration generated
by displacing particles off a lattice, and more generally for any
stochastic particle distribution in $1-d$. It is thus necessary to introduce 
some additional constraint to make this surface contribution 
to the force vanish. 

The previous literature on this model employ top-hat prescriptions
equivalent to Eq.~(\ref{force-symmTH-diff})  to calculate the force,
adding such a constraint.  On the one hand,
Aurell et al. in \cite{aurell_etal} restrict themselves to the study
of an infinite perfect lattice off which only a {\it finite} number 
are initially displaced. In this case the problematic surface 
fluctuation vanishes
for sufficiently large $L$. On the other hand  
\cite{rouet_etal, aurell+fanelli_2002b, miller+rouet_2006,
valageasOSC_1} impose exact symmetry in the displacements 
about some chosen point, which is then taken as the origin of 
the symmetric summation interval. A particle entering (or leaving) at 
one extremity of the interval is then always compensated by one doing 
the same at the other extremity.  

We note that it is only in 
\cite{aurell_etal} that the problem of the infinite system limit 
is actually considered. In the other works the authors do not
discuss this limit explicitly: they consider and study in
practice a finite system, with a prescription for the force
equivalent to Eq.~(\ref{force-symmTH-diff}) where
$2L$ is the system size, i.e., without the explicit
limit $L \rightarrow \infty$.  Symmetry about the 
origin is imposed because this allows one to 
use periodic boundary conditions. Such a {\it finite}
periodic system of period $2L$ is equivalent to
a finite system of size $L$ with reflecting boundary
conditions. The dynamics of such a system is of course
always well defined, for any (finite) initial distribution 
of the points in the box. This does not, however, mean
that this dynamics can be defined in the limit that 
the size of the system is taken to infinity. This is
the question we focus on here, as the definition of
such a limit is essential if a proper analogy is to be
made with the cosmological problem in $3-d$: in this
case 
the gravitational force is well defined in the infinite system limit, 
for a class of statistically translationally invariant distributions 
representing the initial conditions of cosmological 
models\footnote{Numerically one treats, of course, a periodic
system, but it is an {\it infinite} periodic system, i.e., the force
is calculated by summing over the particles in the finite box and all
its (infinite) copies. This is the so-called ``replica
method'', used also widely in equilibrium systems such as the one
component plasma \cite{OCP-review}.  The infinite sum is usually
calculated using the Ewald sum method. To obtain results independent
of the chosen periodic box, the prescription for the force must converge in
the appropriate class of infinite point distributions.}.

The problems with the top-hat prescriptions arise, as we have
seen,  from non-convergent fluctuations at the surface of a
top-hat window, which will be generic in statistically 
translationally invariant point processes. It is thus natural 
to consider smoothing the summation window, and specifically
a prescription for Eq.~(\ref{Force_Reg_secondterm}) such as: 
\begin{equation}
\label{Force1D-int}
F(x) =  g \lim_{\mu \rightarrow 0} \int dy~n(y)~\textrm{sgn}(y-x)~
e^{-\mu|x-y|} \,,
\end{equation}
or, equivalently,
\begin{equation}
\label{Force1D}
F(x) =  g \lim_{\mu \rightarrow 0} 
\sum_{i} \textrm{sgn}(x_{i}-x) 
e^{- \mu \vert x_{i}- x \vert} \,,
\end{equation}
where the sum runs over all particles in the (infinite) 
distribution. Rather than a smoothing of the summation window,
this can be interpreted more physically in terms of
the screening of the gravitational interaction, i.e.,
the pair force law of Eq.~(\ref{Familiar_pair_force}) is 
replaced by
\begin{equation}
  \label{Screening}
  f_{\mu}(x) = -g\,\textrm{sgn} (x) \, e^{-\mu\,\vert  x \vert},
\end{equation}
and the gravitational force in the infinite system limit is defined 
as that obtained when the screening length is taken to infinity,
{\it after} the infinite system is taken\footnote{Although we will not use
the interparticle potential in our calculations, we note 
that $f_\mu (x)=-d\phi_\mu/dx$ where 
$\phi_\mu(x)=-g e^{-\mu \vert x \vert}/\mu$ is the solution
of $\frac{d^{2}\phi_\mu}{dx^{2}} - \mu^2 \phi_\mu=2g \delta_{D} (x)$.} (see
Fig. 2).
This treatment is borrowed from the class of infrared problems 
well known in quantum field theory. The standard procedure of handling
infrared divergences is to apply an infrared regularization, to solve
the regularized problem, and to remove the regularization at the end 
of the calculation, perhaps involving a renormalization.

\begin{figure}[!htb]
\begin{center}
 \includegraphics[width=9cm]{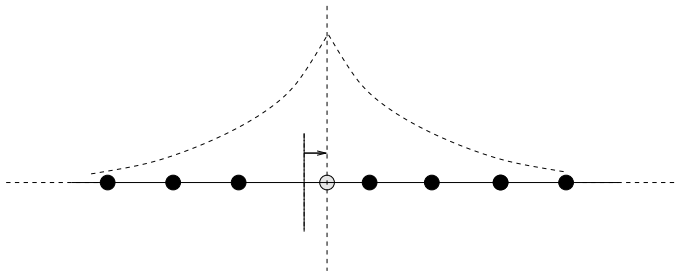}
\label{screening}
\end{center}
\caption{Schematic representation of the smooth screening of the force
 (or, equivalently, summation window).}
\end{figure}

For the case of a single particle displaced off a perfect lattice 
discussed above it is simple to calculate the force using 
Eq.~(\ref{Force1D-int}).
Denoting the lattice spacing by $\ell$, and
the displacement by $u$, we have 
\begin{equation}
\label{F_basic_conf}
 F(u) = g  \lim_{\mu \rightarrow 0} \sum_{n\neq 0} \textrm{sgn} (n\ell-u) 
e^{- \mu \vert n\ell-u \vert}.
\end{equation}
For $|u| \leq \ell$ 
the sum gives
\begin{equation}
2\, \sinh(\mu u )\, \Big( \sum_{n > 0} e^{-\mu n\ell} \Big).
\end{equation}
Expanding this in powers of $\mu$ we obtain
\begin{equation}
 F_{\mu}(u) = \frac{2 g u}{\ell} + O(\mu).
\end{equation}
Taking the limit $\mu \rightarrow 0$ gives 
Eq.~(\ref{force-single-particle}), i.e., 
the result obtained using the top-hat prescription Eq.~(\ref{force-symmTH-diff}).  
The equivalence of the two prescriptions can likewise be shown to apply
when displacements are applied to a {\it finite} number of particles on the 
lattice (which leave the forces unchanged, and equal to
Eq.~(\ref{force-single-particle}), if there are no crossings).
Thus the only difference between the prescriptions is how they 
treat the contribution from particles at arbitrarily large distances 
when the infinite system limit is taken.
 
We will show rigorously in the next section 
that, for a class of infinite perturbed lattices in which
particles do not cross, the prescription Eq.~(\ref{Force1D-int}) 
simply removes the problematic surface contribution present
in the top-hat prescriptions (without applying any additional
constraint of symmetry). This gives a force on each particle 
equal to Eq.~(\ref{force-single-particle}) where $u$ is the 
displacement of the particle, the only difference with respect 
to the case of a finite number of displaced particles being 
that the origin of this displacement may be redefined by a 
net translation of the whole system induced by the infinite 
displacements. The force felt by each particle is thus equivalent 
to that exerted by an inverted harmonic oscillator about
an (unstable) equilibrium point. We note that this
expression for the force is in fact what one would expect
from a naive generalization of the analagous results in $3-d$.
In the latter case it can be shown \cite{gabrielli_06} that 
the force on a single particle displaced off an infinite 
lattice by a vector ${\bf u}$ is, to linear order 
in $|{\bf u}|$,  simply
\be
{\bf F} ({\bf u}) = 4\pi G \rho_0 {\bf u}/3 \;.
\label{force-single-particle-3d}
\ee
This force is simply that which is inferred, by Gauss's law,  
as due to a uniform background of mass density -$\rho_0$  
(i.e. due to the mass of such a background contained in 
a sphere of radius $|{\bf u}|$). The $1-d$ result 
is exactly analogous, as $2n_0 |u|$ is simply the mass 
inside the interval of ``radius''  $|u|$. While this
result is valid, in $3-d$, only at linear order and 
for the case of a single displaced particles, 
it is exactly valid in $1-d$ in absence of particle 
crossings and for a broad class of displacement
statistics. The reason is simply that in $1-d$ the force
on a particle is unaffected by displacements of other 
particles, unless the latter cross the considered 
particle.

\section{Forces in infinite perturbed lattices}
\label{Forces in infinite perturbed lattices}

In this section we calculate, using the definition Eq.~(\ref{Force1D}),
the gravitational force on particles in a class of infinite perturbed 
lattices. To do this we describe these point distributions
as generated by a stochastic process in which the particles
are displaced\footnote{For an introduction to the formalism of 
stochastic point processes \textit{i.e.} stochastic spatial 
distributions of point-particles with identical mass,
see, e.g., \cite{book}.}. The force on a particle (or the
force field at a point in space) is then itself a stochastic 
variable, taking a different value in each realization of 
the point process, and the question of its definedness 
can be cast in terms of the existence of the probability 
distribution function (PDF) of the force. We thus calculate here
the PDF of the force on a particle with a given displacement $u$, 
in the ensemble of realizations of the displacements of the other 
particles. The result is that, for 
the class of stochastic displacement fields in which displacements
are such that particles do not cross, this force PDF
becomes simply a Dirac delta function. This gives the
anticipated result, that the only force which results
is that due to the particle's own displacement 
given by Eq.~(\ref{force-single-particle}),
modulo an additional term describing a contribution 
from the coherent displacement of  the whole infinite 
lattice if the average displacement is non-zero.


\subsection{Stochastic perturbed lattices}
Let us consider first an infinite $1-d$ regular chain of unitary 
mass particles with lattice spacing $\ell>0$, i.e., the position of 
the $n^{th}$ particle is $X_{n} = n\ell$,  
and the microscopic number density can be written as 
\begin{equation}
 n_{in}(x) = \sum_{n=-\infty}^{+\infty} \delta_{D} (x-n \ell).
\end{equation}
We now apply a stochastic displacement field $\{U_{n}\}$ to this
system, in which the displacement $U_{n}$ is applied to the generic
$n^{th}$ particle with $n\in \mathbb{Z}$.  Let us call $\{u_n\}$ the
single realization of the stochastic field $\{U_{n}\}$. The
corresponding realization of the point process thus has 
microscopic
number density
\begin{equation}
 n(x) = \sum_{n=-\infty}^{+\infty} \delta_{D} (x-n\ell-u_{n})\,.
\end{equation}
This displacement field is completely characterized by the joint
displacement PDF ${\cal P}(\{ u_{n}\})$ where $\{ u_{n}\}$ is the set
of all particle displacements with $n \in \mathbb{Z}$. We will further
assume that this stochastic process is statistically translationally
invariant, i.e. ${\cal P}(\{ u_{n}\})={\cal P}(\{ u_{n+l}\})$ for any
integer $l$.  This implies in particular that the one displacement PDF
(for the displacement applied to a single particle) is independent of
the position of that particle, i.e., the function
\begin{equation}
 p_m(u)\equiv\int \prod_n du_n{\cal P}(\{ u_{n}\})\delta_D(u-u_m) 
\end{equation}
is independent of $m$, i.e. $p_m(u)=p(u)$. Moreover the joint
two-displacement PDF
\[q_{nm}(u,v)=\int \prod_n du_n{\cal P}(\{ u_{n}\})
\delta_D(u-u_m)\delta_D(v-u_n)\]
depends parametrically on the lattice positions $n,m$ only through their
relative distance $(m-n)$.
\subsection{Mean value and variance of the total force}

Let us denote in general by $F_{\mu}(x_0)$ the total gravitational force, 
with finite screening $\mu$, acting on the particle at $x_{0}$ and due
to all the other particles placed at $x_n$:
\begin{equation}
 F_{\mu}(x_0) =  g \sum_{n\neq 0} \textrm{sgn}(x_{n}-x_{0}) 
e^{- \mu \vert x_{n}-x_{0} \vert}\,.
\label{F1}
\end{equation}
Writing now $x_n=n\ell+u_n$ in Eq.~(\ref{F1}), we can write
the total screened force on the particle at $x_0=u_0$ in a perturbed lattice
for a given realization of the displacement field:  
\begin{equation}
 F_{\mu}(u_0) =  g \sum_{n\neq 0} \textrm{sgn}(n\ell+u_{n}-u_{0}) 
e^{- \mu \vert n\ell+ u_{n}-u_{0} \vert}.
\end{equation}
Note that, given the assumed statistical translational invariance of
the field $\{U_n\}$ the statistical properties of the force are the
same for all particles in the system.  If, further, we assume now that
the displacements from the lattice are such that {\it particles do not
cross}, i.e.  $\textrm{sgn}(n\ell+u_{n}-u_{0})=\textrm{sgn}(n)$ for
$n\ne 0$, this can be written as
\begin{equation}
\label{F_screen}
F_{\mu}(u_0) = g \sum_{n=1}^\infty e^{-\mu n \ell}f_{n}, 
\end{equation}
where we define for, $n \geq 1$,
\begin{equation}
\label{def-fn} 
f_{n}\equiv f_{n}(\mu)=e^{-\mu (u_{n}-u_{0})}-e^{-\mu (u_{0}-u_{-n})}.
\nonumber
\end{equation}

We now take the average of Eq.~(\ref{F_screen}) over all realizations
of the displacements of all particles, except the chosen one $u_0$,
which we consider as fixed. We denote this conditional average as
$\left<\cdot\right>_0$,  while we use $\left<\cdot\right>$ for the
unconditional average.  In order to do this we need the conditional PDF of
$U_n$ to $U_0$, which by definition of conditional probability is
\begin{equation}
P_n(u;u_0)=\frac{q_{n0}(u,u_0)}{p(u_0)}\,.
\label{P_n}
\end{equation}
By using this function we can write
\begin{equation} 
\la f_n(\mu)\ra_0=e^{\mu u_{0}}
\tilde{P}_n (\mu;u_0) - e^{-\mu u_{0}} \tilde{P}_{-n} (-\mu;u_0) 
\label{fmu}
\end{equation}
and therefore
\begin{equation}
\label{F_screen_mean}
\big< F_{\mu}(u_0) \big >_0 = g \sum_{n=1}^\infty \left[e^{\mu u_{0}}
\tilde{P}_n (\mu;u_0) - e^{-\mu u_{0}} \tilde{P}_{-n} (-\mu;u_0) \right]
e^{-\mu n \ell}
\end{equation}
where we have defined 
\begin{eqnarray}
\label{pmu}
\tilde{P}_n (\mu;u_0) &=&\int_{-\infty}^{\infty} du P_n(u;u_0) e^{-\mu u}, \\
&=& \sum_{k=0}^\infty \frac{(-\mu)^k \left<U_n^k\right>_0}{k!}. \nonumber
\end{eqnarray}
The latter equality is valid when all the moments $\big <U^k_n\big>_0$
of $P_n(u;u_0)$ are finite. Note that, given the assumption that
particles do not cross, it follows from the definition (\ref{P_n}) 
that $q_{n0}(u,u_0)=0$ for $u + n\lessgtr u_0$ respectively 
for $n\gtrless 0$. Therefore $P_n(u;u_0)$ is always zero for some
sufficiently negative $u_0$ dependent value of $u$ if 
$n>0$, and likewise for sufficiently positive values if
$n<0$. This
ensures that the integral in Eq.~(\ref{pmu}) is indeed finite.

In order to study the behavior of Eq.~(\ref{F_screen_mean}) for
$\mu\to 0$, we will assume that
\be
\label{decay-correlations}
q_{nm}(u,v)\stackrel{|n-m|\to \infty}{\longrightarrow}p(u)p(v)\,.
\ee 
This corresponds to the assumption that the displacement field is
a well defined stochastic field, which requires (see e.g. \cite{book})
that the two-displacement correlations vanish as the spatial 
separation diverges. We will discuss in the next section the
restriction this corresponds to on the large scale behaviour of
the density perturbations, which is of particular relevance when
one considers the analogy to $3-d$ cosmological simulations.

Assuming Eq.~(\ref{decay-correlations}) we can write
\[P_n(u;u_0)=p(u)+r_n(u;u_0)\,,\]
where $r_n(u;u_0)$ is a function vanishing for $|n|\to \infty$ and with
zero integral over $u$ for any $n$. As a consequence 
\be 
\tilde{P}_n
(\mu;u_0)=\tilde p(\mu)+\tilde{r}_n(\mu;u_0)\,,
\label{pmu2}
\ee 
where we used the definition analogous to Eq.~(\ref{pmu}) for 
$\tilde p(\mu)$ and 
$\tilde r_n(\mu;u_0)$, and the latter vanishes for $\mu\to 0$ and/or
$n\to \infty$. If we now suppose that both $\left<U\right>$ and
$\left<U_n\right>_0$ are finite, with evidently
$\left<U_n\right>_0\to\left<U\right>$ for $n\to\infty$, we can write
at lower order: \bea
\label{pmu-exp}
&&\tilde p(\mu)=1-\mu\la U\ra+o(\mu),\\ &&\tilde{r}_n(\mu;u_0)=\mu(\la
U\ra-\la U_n\ra_0)+o(\mu)\,.\nonumber 
\eea 
It is now simple, by substituting Eqs.~(\ref{pmu2}) and
(\ref{pmu-exp}) into Eq.~(\ref{F_screen_mean}), to show that, if
$\big( \la U\ra-\la U_n\ra_0 \big)$ decays in $n$ as a negative power
law or faster, we have
\begin{equation} 
\label{force-mean}
\la F (u_0) \ra_0 \equiv \lim_{\mu \rightarrow 0} \big< F_{\mu}(u_0)
\big >_0 = 2gn_0(u_0-\left<U\right>)\,.
\end{equation}

We will now show that both for uncorrelated displacements, and then
more generally for correlated displacements with decaying
correlations, this average force is in fact the exact force in every
realization. We do so by simply showing that
\begin{equation} 
\label{zero-dispersion}
\lim_{\mu \rightarrow 0} \left[ \la F_{\mu}^2(u_0) \ra_0 
-\la F_\mu(u_0) \ra_0^2 \right]= 0 \,.
\end{equation}
This implies that the variance of the conditional PDF of the total
force $F$ acting on the particle at $u_0$ vanishes, i.e., this
PDF is a Dirac delta function at the average value given by
Eq.~(\ref{force-mean}). Compared to the simple case of a single
displaced particle we analysed above, the only effect of the (infinite
number of) other displacements is to possibly shift the centre 
of mass of the whole (infinite) distribution with respect to which the
displacement of the single particle is defined.  

In order to show
Eq.~(\ref{zero-dispersion}) we note first that the second conditional
moment of $F$ may be written
\begin{eqnarray}
 \la F_\mu^2 (u_0)\ra_0 & = & g^2 \sum_{n,m}^{1,\infty}e^{-\mu (n+m)\ell}
\la f_{n}\,f_{m} \ra_0 \nonumber \\
 & = & \la F_\mu (u_0)\ra_0^2 +   g^2 \sum_{n=1}^\infty e^{-2\mu n \ell} A_n (\mu) \nonumber \\
&&+ g^2 \sum_{n,m}^{1,\infty}{}^{'} e^{-\mu (n+m)\ell} B_{nm} (\mu), 
\label{2nd-moment-explicit}
\end{eqnarray}
with
\begin{eqnarray}
\label{2-mom}
A_n (\mu) &=& \la f_n^2\ra_0 - \la f_n\ra_0^2, \\
B_{nm} (\mu) &=& \la f_n f_m\ra_0 - \la f_n\ra_0\la f_m\ra_0 \quad 
(m \neq n),
\nonumber  
\end{eqnarray}
and where $\sum_{n,m}'$ as usual indicates the sum over $m$ and $n$
with the exception of the $n=m$ terms. To prove
Eq.~(\ref{zero-dispersion}) it is sufficient to show that the last two
terms in Eq.~(\ref{2nd-moment-explicit}) go continuously to zero as
$\mu$ does so.

\subsection{Lattice with uncorrelated displacements}

We consider first the case that the displacements are uncorrelated
and identically distributed, i.e., 
\begin{equation}
{\cal P}(\{ u_{n}\})=\prod_{n=-\infty}^{+\infty} p(u_n).
\end{equation}
We refer to this as a ``shuffled lattice'' configuration (following
\cite{book}). In this case conditional and unconditional averages
coincide.  Given the assumption that the displacements do not make
particles cross, we must have that $p(u)=0$ for $|u| > \ell/2$,
implying that all the moments of $p(u)$ are necessarily finite.

In this case the $u_n$ are statistically independent 
and identically distributed random variables. Given the
definition Eq.~(\ref{def-fn}), it follows that the 
$f_n$ also have this property, i.e., 
\begin{equation}
\la f_n f_m\ra=\la f_n\ra \la f_m\ra \,,
\end{equation}
and thus that $B_{nm} (\mu)=0$. Further $A_n(\mu)$ is 
independent of $n$ and can be expressed explicitly as 
\begin{equation}
\label{An-uncorrelated}
A_n (\mu)= 
e^{2\mu u_{0}} \big[\tilde{p} (2\mu) - \tilde{p}^2 (\mu) \big] 
- e^{-2\mu u_{0}} \big[\tilde{p} (-2\mu) - \tilde{p}^2 (-\mu) \big]. 
\end{equation}
Expanding this expression in $\mu$ about $\mu=0$, we
find that the leading non-vanishing term is at order $\mu^2$.
The desired result, Eq.~(\ref{zero-dispersion}), follows as
\[\sum_{n=1}^{\infty} e^{-2\mu nl}=\frac{e^{-2\mu l}}{1-e^{-2\mu l}}
=O(\mu^{-1})\;\;\;\mbox{for }\mu\to 0\,,\]
where $O(\mu^l)$ means as usual a term of order $l$ in $\mu$.

\subsection{Lattice with correlated displacements}

We now consider the case where the displacements are non-trivially
correlated.  In order to calculate $A_n(\mu)$ and $B_{nm}(\mu)$ we
need both the conditional single displacement PDF
$P_n(u;u_0)$ and the conditional two-displacement PDF
$Q_{nm}(u,v;u_0)$, both conditioned to the fixed value $u_0$ of the
stochastic displacement $U_0$. The function $Q_{nm}(u,v;u_0)$ is 
defined by the rules of conditional probability as
\[Q_{nm}(u,v;u_0)=\frac{s_{nm0}(u,v,u_0)}{p(u_0)},\]
where $s_{nml}(u,v,w)$ is the joint three displacement PDF of having
the three displacements $u,v,w$ respectively at the lattice sites
$n,m,l$.

Let us start from the evaluation of $A_n(\mu)$.
From its definition it is simple to show that 
\bea
\la f^2_n(\mu)\ra_0 = &&e^{2\mu u_0}\tilde P_n(2\mu;u_0)+
e^{-2\mu u_0}\tilde P_{-n}(-2\mu;u_0)\nonumber\\
&&-2\tilde Q_{n\,-n}(\mu,-\mu;u_0),
\label{f2mu}
\eea
where 
\[\tilde Q_{nm}(\mu,\nu;u_0)=\int\int_{-\infty}^{+\infty}du\,dv\,
Q_{nm}(u,v;u_0) e^{-(\mu u+\nu v)}.\] In order to study the limit
$\mu\to 0$ we have to expand $\tilde P_n(\mu;u_0)$ and $\tilde
Q_{nm}(\mu,\pm\mu;u_0)$ in powers of $\mu$. Assuming
that at least the first two moments of the displacement 
statistics are finite, we can 
write
\bea
\nonumber
\tilde P_n(\mu;u_0) &=& 1-\mu\la U_n\ra_0+\frac{\mu^2}{2}\la U_n^2\ra_0
+o(\mu^2),\\
\tilde Q_{nm}(\mu,\pm\mu;u_0) &=& 1 -\mu\left(\la U_n\ra_0\pm\la U_m\ra_0\right)
+ \frac{\mu^2}{2}\left(\la U_n^2\ra_0\right.\nonumber\\
&&\left.+\la U_m^2\ra_0\pm\la U_nU_m\ra_0\right)+o(\mu^2).\nonumber\\
\label{p-q-mu}
\eea
Using this result and Eqs.~(\ref{fmu}) and (\ref{f2mu})
in the definition (\ref{2-mom}) of $A_n(\mu)$, it is 
simple to show that
\bea
\label{a-mu-as}
A_n(\mu)=&&\mu^2\left[e^{2\mu u_0}\left(\la U_n^2\ra_0-\la U_n\ra_0^2\right)
\right.\\
&&+e^{-2\mu u_0}\left(\la U_{-n}^2\ra_0-\la U_{-n}\ra_0^2\right)\nonumber\\
&&\left.+2\left(\la U_nU_{-n}\ra_0-\la U_{n}\ra_0\la U_{-n}\ra_0\right)
\right]+o(\mu^2)\,.\nonumber
\eea
Note that for $|n|\to\infty$ we have $\la U_n\ra_0\to \la U\ra$, 
$\la U_n^2\ra_0\to \la U^2\ra$ and $\la U_nU_{-n}\ra_0\to \la U\ra^2$.
Therefore we can write
\[A_n(\mu)\stackrel{n\to \infty}{\longrightarrow}\mu^2(\la U^2\ra-\la U\ra^2)
(e^{2\mu u_0}+e^{-2\mu u_0})\,,\]
where we have used the fact that, as the coefficients of the higher order 
contributions in $\mu$ to $A_n(\mu)$ are non-diverging, they can be 
neglected.
This is sufficient to conclude that 
\be
\sum_{n=1}^\infty e^{-\mu n}A_n(\mu)=O(\mu)\,,
\label{A-n-as}
\ee
where $O(\mu^l)$ as usual means a term of order $\mu^l$, and therefore
the sum vanishes as $\mu$ for $\mu\to 0$.

Let us now move to analyze the last sum in Eq.~(\ref{2nd-moment-explicit}).
We study the behavior of $B_{nm}(\mu)$ as defined by Eq.~(\ref{2-mom}).
It is simple to show that 
\bea
\la f_nf_m\ra_0 &=& e^{-2\mu u_0}\tilde Q_{nm}(\mu,\mu;u_0)+e^{2\mu u_0}
\tilde Q_{-n\,-m}(-\mu,-\mu;u_0)\nonumber\\
 &{}& - \tilde Q_{n\,-m}(\mu,-\mu;u_0)-\tilde Q_{-nm}(-\mu,\mu;u_0).
\label{fn-fm-av}
\eea
Using this equation together with Eqs.~(\ref{2-mom}),(\ref{fmu})
and (\ref{p-q-mu}), we can write
\bea
B_{nm}(\mu) &=& \mu^2 [e^{-2\mu u_0}g(n,m;u_0)+e^{2\mu u_0}g(-n,-m;u_0)
\nonumber\\
&{}& -g(n,-m;u_0)-g(-n,m;u_0)]+o(\mu^2),\nonumber \\
\label{b-m-n-as}
\eea
where we have called 
\[g(n,m;u_0)=\la U_nU_m \ra_0-\la U_n\ra_0\la U_m \ra_0\,,\]
i.e., the conditional displacement covariance matrix.
Since this is a ``conditional'' correlation it does not depend
simply on $n-m$, but on both $n$ and $m$ in a non-trivial way.
However for both $|n|,|m|\to\infty$ the conditional averages coincide with
the unconditional ones and therefore we can write
\be
g(n,m;u_0)=c(|n-m|)[1+h(n,m;u_0)]\,,
\label{g-vs-c}
\ee
where $c(|n-m|)=\la U_nU_m \ra-\la U\ra^2$ is the unconditional
displacement covariance matrix, and $h(n,m;u_0)\to 0$ for 
$|n|,|m|\to \infty$.
In order to analyze the asymptotic behavior for small $\mu$ of
\be
I(\mu)\equiv \sum_{n,m}^{1,\infty}{}^{'}e^{-\mu(n+m)}B_{nm}(\mu),
\label{I-B}
\ee 
it is sufficient to study the behavior of the sum coming from the
first term (or equivalently the second) of $B_{nm}(\mu)$ in
Eq.~(\ref{b-m-n-as}) as it is the most slowly convergent one, i.e.,
basically to study the following sum:
\[J(\mu)=\sum_{n,m}^{1,\infty}{}^{'}e^{-\mu(n+m)}g(n,m;u_0)\,.\]
Since $h(n,m;u_0)\to 0$ for $|n|,|m|\to \infty$, the small $\mu$
scaling behavior of $J(\mu)$ is the same if we replace $g(n,m;u_0)$
by $c(|n-m|)$: \be J(\mu)\simeq
\sum_{n,m}^{1,\infty}{}^{'}e^{-\mu(n+m)}c(|n-m|)\,.
\label{I-c}
\ee
This can be also shown by the following argument:
assuming that $h(n,m;u_0)$ is bounded, say $|h(n,m;u_0)|\le A$, we can write
\[
\begin{array}{l}
\left|J(\mu)\right|\le \sum_{n,m}^{1,\infty}{}^{'}e^{-\mu(n+m)}|g(n,m;u_0)|\\
\le(1+A) \sum_{n,m}^{1,\infty}{}^{'}e^{-\mu(n+m)}|c(|n-m|)|\,.
\end{array}
\]
Therefore the convergence to zero of $\mu^2$ times the right-hand side
of Eq.~(\ref{I-c}) is a sufficient condition to have the variance of
$F$ to vanish for $\mu\to 0$.   

Let us now analyze the right-hand side of Eq.~(\ref{I-c}). We can
write 
\bea
&&\sum_{n,m}^{1,\infty}{}^{'}e^{-\mu(n+m)}c(|n-m|)\nonumber\\
&&=\sum_{n,m}^{1,\infty}
e^{-\mu(n+m)}c(|n-m|)-c(0)\frac{1}{e^{2\mu}-1},
\label{sum-sum}
\eea
where $c(0)$ is the
single displacement variance.  Note that the second term is of order
$\mu^{-1}$ at small $\mu$ and therefore gives rise to a term 
at linear order in $\mu$ 
in Eq.~(\ref{I-B}). Let us introduce the Fourier transform $\tilde c(k)$
of $c(n)$, defined by
\[c(n)=\int_{-\pi}^{\pi}\frac{dk}{2\pi}\tilde c(k)e^{ikn}\,.\]
Using this in the right-hand side of Eq.~(\ref{sum-sum}) we get
\bea
\label{FT-c}
&&\sum_{n,m}^{1,\infty}e^{-\mu(n+m)}c(|n-m|)\\ &&=\int_{-\pi}^{\pi}
\frac{dk}{2\pi}\tilde c(k)\frac{1}{e^{2\mu}+1-2e^\mu\cos k}\,.
\nonumber 
\eea 
The small $\mu$ limit of this integral is dominated by the behavior at
small $k$ of the integrand.  In this limit the following approximation
holds $(e^{2\mu}+1-2e^\mu\cos k)\simeq (\mu^2+k^2)$.  Let us also
assume that $c(n)\sim n^{-\alpha}$ at large $n$ (with in general
$\alpha>0$)\footnote{The case of a decay faster than any power,
e.g. exponential decay, can be included for $\alpha\to\infty$.}  which
implies at small $|k|$ $\tilde c(k)\sim |k|^{\alpha-1}$ for
$0<\alpha\le 1$ (with logarithmic corrections for $\alpha=1$) and
$\tilde c(k)\sim |k|^{\beta}$ with $\beta\ge 0$ for $\alpha>1$.
Therefore the small $\mu$ behavior of Eq.~(\ref{FT-c}) is the same
as that of the simple integral 
\be 
\int_{-\pi}^{\pi} \frac{dk}{2\pi}
\frac{\tilde c(k)}{\mu^2+k^2}\sim\left\{
  \begin{array}{ll}
              \mu^{\alpha-2}&\mbox{for }0<\alpha\le 1,\\
              \mu^{\beta-1}&\mbox{for }\alpha>1.
  \end{array}
\right.
\label{B-asym1}
\ee 
Taking also into account the second term in Eq.~(\ref{sum-sum}), we
can therefore conclude that \be
\sum_{n,m}^{1,\infty}{}^{'}B_{nm}(\mu)e^{-(n+m)\mu} \sim \left\{
   \begin{array}{ll}
              \mu^{\alpha}&\mbox{for }0<\alpha<1,\\
              \mu&\mbox{for }\alpha\ge 1\,.
   \end{array}
\right.
\label{B-asym2}
\ee 
This, together with the results for the first sum in
Eq.~(\ref{2nd-moment-explicit}), it follows that
at small $\mu$
\be
\left<F^2_\mu(u_0)\right>_0-\left<F_\mu(u_0)\right>_0^2 \sim \left\{
   \begin{array}{ll}
           \mu^{\alpha}&\mbox{for }0<\alpha<1,\\
           \mu&\mbox{for }\alpha\ge 1,
   \end{array}
\right.
\label{F2}
\ee 
i.e. it vanishes in the $\mu\to 0$ limit and the PDF of the total
force acting on a particle displaced by $u_0$ from its lattice position
is $W(F;u_0)=\delta[F-2g(u_0-\la U\ra)]$. In other words, even in the
case of spatially correlated displacements, the total force acting on
a particle is a deterministic quantity equal to $2g(u_0-\la U\ra)$
with no fluctuations. 
This value depends only on the
displacement of the particle on which we are calculating the force and
not on the displacements of other particles 
as it does in $3-d$ \cite{gabrielli_06}.


\section{Dynamics of 1d gravitational systems} 
\label{Dynamics of 1d gravitational systems} 

In the previous section we have shown the prescription 
Eq.~(\ref{Force1D-int}) for the $1-d$ gravitational force 
to give a well defined result in a class of infinite 
displaced lattice distributions. This result can be used 
in the construction of different toy models, through 
different prescriptions for the dynamics associated to 
these forces. In this section we discuss two such models, 
analogous to the $3-d$ cases of gravitational clustering 
in an infinite static or expanding universe, respectively. 
In the last subsection we discuss in detail the relation of 
these models to previous treatments of such models in the 
literature. 

As motivation let us first comment on the reason for our
interest in the case of perturbed lattices: in $3-d$ 
cosmological $N$-body simulations precisely such 
configurations are used as initial conditions. The reason
is that by displacing particles from a lattice in this
way, one can represent accurately, at sufficiently large 
scales, low-amplitude density perturbations about 
uniformity with a desired power spectrum $P(k)$ (for a detailed 
discussion see e.g. \cite{book} or \cite{discreteness1_mjbm}).
This algorithm is strictly valid in the limit of very
small relative displacements of particles, so that the
assumption that particles do not cross in our derivation
is a reasonable one (although not, as we will discuss in
our conclusions, rigorously valid). The further 
assumption Eq.~(\ref{decay-correlations}) we have made, on the 
decay of correlations, corresponds, also to a reasonable 
restriction on the class of initial power spectra.
Indeed it can be shown easily that it corresponds,
in $d$ dimensions, to the assumption that $P(k)/k^2$ 
be integrable at $k=0$. In $3-d$ this corresponds to
$P(k\rightarrow 0) \sim k^n$ with $n>-1$, which is
strictly satisfied in typical cosmological models
which are characterised by an exponent $n=1$ at
asymptotically small $k$.

\subsection{Toy models: static}
The simplest such model is the conservative
Newtonian dynamics associated to the derived force law, 
i.e., with equation of motion
\begin{equation}
\label{hamiltonian}
\ddot{x}_i=F_i (\{ x_j, j=0..\infty\}, t),
\end{equation}
where $F_i$ is the gravitational force on the 
$i$-th particle of the distribution, with position $x_i$ at time $t$ (and
dots denote derivatives with respect to $t$), calculated
using the prescription Eq.~(\ref{Force1D}), i.e., 
\begin{equation}
\label{sheets-infinite-static}
\ddot{x}_i = -g
\lim_{\mu \rightarrow 0} \sum_{j\neq i}
\textrm{sgn}(x_i - x_j) e^{-\mu \vert { x}_i - {x}_j \vert}. 
\end{equation}
 We have shown
that, for the case of an infinite lattice subjected to
displacements which (i) do not make the particles cross,
and (ii) satisfy Eq.~(\ref{decay-correlations}), 
the force on the right-hand side is simply given 
deterministically as 
proportional to the particle's displacement
(when $\left<U\right>$, the average displacement, is zero).
Denoting then the displacements of the $i$-th particle 
by $u_i$, i.e. $x_i=i\ell+u_i$, the equation of motion is 
therefore 
\begin{equation}
\label{inverted-oscillator}
\ddot{u}_{i}(t) = 2gn_0 u_{i}(t) \,,
\end{equation}
i.e., simply that of an inverted harmonic oscillator.  The same
equation is valid in the case that $\left<U\right>\neq0$ if
we define $x_i=i\ell+\left<U\right>+u_i$. This equation of
motion is valid, of course, only as long as the non-crossing 
condition is satisfied. While it is
in principle straightforward to generalize our calculation of the
force to incorporate the effects of a finite number of crossings, it
is much more convenient to make use of the following fact, which we
recalled above: particles crossings in $1-d$ are equivalent, up to
exchange of particle labels, to elastic collisions between particles,
in which velocities are exchanged.  This means that if we are
interested in properties of the model which do not depend on particle
labels, the model of $1-d$ self-gravitating particles is {\it
equivalent} to a model in which particles bounce elastically. In this
case the particles displacements from their original lattice sites are
{\it at all times} such that there is no crossing of particles, and
Eq.~(\ref{inverted-oscillator}) remains valid, except exactly at
``collisions''. The dynamics of this model is therefore equivalent to
that of an infinite set of inverted harmonic oscillators centred on
the sites of a perfect lattice which bounce elastically, exchanging
velocities, when they collide.  As in the finite ``sheet model'' the
equation of motion may be integrated exactly. Defining, for
convenience, time in units of the characteristic ``dynamical'' time
$\tau_{dyn} = 1/\sqrt{2gn_0}$, the evolution between collisions is
given exactly by
\begin{eqnarray}
\label{eqns-integrated-static-u}
u_{i}(t_0+t) = u_{i}(t_0) \, \cosh t + v_{i}(t_0) \, \sinh t,\\
\label{eqns-integrated-static-v}
v_{i}(t_0+ t) = u_{i}(t_0) \, \sinh t + v_{i}(t_0) \, \cosh t,
\end{eqnarray}
where $u_i(t_0)$ [$v_i(t_0)$] is the position (velocity) after the preceeding
collision. The solution of the dynamics requires simply the
determination of the next crossing time, which involves 
the solution of a quadratic equation (in $e^{t}$), followed
by an appropriate updating of the velocities of the colliding
particles.

\subsection{Toy models: expanding}

The model we have just discussed is the $1-d$ analogy for
the problem of gravitational clustering in an infinite
static universe, with equations of motion
\begin{equation}
\label{3d-equations-physical}
\ddot{\bf r}_i = -Gm {\sum}^{J}_{j\neq i}
\frac{{\bf r}_i - {\bf r}_j}{\vert {\bf r}_i - {\bf r}_j \vert^3}\,,
\end{equation}
for identical particles of mass $m$. We use the superscript $J$ 
on the sum to indicate that the sum is calculated using the
Jeans swindle. As we have discussed this ``swindle'' in 
$3-d$ can be implemented by summing symmetrically 
about the point $i$ either in a top-hat (i.e. sphere) or
using the limiting procedure with a screening.  

The equations of motion for particles in an infinite {\it expanding}
$3-d$ universe are usually written in the form 
\begin{equation}
\label{3d-equations}
\ddot{\bf x}_i +
2H \dot {\bf x}_i = -\frac{Gm}{a^3}{\sum}^J
\frac{{\bf x}_i - {\bf x}_j}{\vert {\bf x}_i - {\bf x}_j \vert^3} \,,
\end{equation}
where ${\bf x}_i$ are the so-called {\it comoving coordinates} 
of the particles, $H(t)={\dot a}/{a}$ is the Hubble ``constant'',
and $a(t)$ is the scale factor which is a solution of the 
equations 
\begin{eqnarray}
H^2 &=& \left(\frac{\dot{a}}{a} \right)^2=\frac{8 \pi G}{3a^3}\rho_0 +
 \frac{C}{a^2} \,, 
\label{hubble-law} \\
\frac{\ddot{a}}{a} &=& -\frac{4\pi G}{3a^3}\rho_0 \,,
\label{ddot-a} 
\end{eqnarray}
where $\rho_0$ is the mean mass density when $a=1$, 
and $C$ is a constant of integration\footnote{$C=0$ corresponds to the
flat Einstein de Sitter universe, $C>0$ to a closed universe,
and $C<0$ to an open universe. In the Newtonian derivation of
these equations, given below, $C$ can be expressed in terms
of the physical particle velocities at some initial time.}. 

Note that these equations can be derived entirely in a
Newtonian framework, and correspond simply to a different
regularisation of the infinite system limit than that 
employed in the Jeans swindle: instead of discarding
the effect of the mean mass density, the force is
regularised so that the mean density sources a
homologous expansion (or contraction) of the whole
system. This corresponds to taking equations of
motion 
\begin{equation}
\label{3d-equations-expanding}
\ddot{\bf r}_i= -Gm \lim_{R\rightarrow \infty}
\sum_{j\neq i, \vert {\bf r}_j \vert <R}
\frac{{\bf r}_i - {\bf r}_j}{\vert {\bf r}_i - {\bf r}_j \vert^3} \,,
\end{equation}
i.e. with the sum for the force calculated by summing 
symmetrically {\it about a chosen origin}. Dividing the sum
into a term due to the mean mass density and a term due
to fluctuations about this density, this may be written
as
\begin{equation}
\label{3d-equations-physical-bgd}
\ddot{\bf r}_i= - \frac{4 \pi G \rho}{3} {\bf r}_i - Gm {\sum}^J
\frac{{\bf r}_i - {\bf r}_j}{\vert {\bf r}_i - {\bf r}_j \vert^3}
\,,
\end{equation}
Neglecting the second term (i.e. taking only the force due
to the mean density) gives an equation of motion admitting 
solutions of the form ${\bf r}_i(t)= a(t) {\bf r}_i(t_0)$, 
with $a(t)$ satisfying Eqs. (\ref{hubble-law}) and (\ref{ddot-a}). 
Changing to comoving coordinates defined by 
${\bf r}_i=a(t){\bf x}_i$ in Eq.~(\ref{3d-equations-expanding}) 
[or in Eq.~(\ref{3d-equations-physical-bgd})],
and using Eq.~(\ref{ddot-a}), then gives Eq.~(\ref{3d-equations}).

Note that setting $a(t)=1$ 
in Eq.~(\ref{3d-equations}) gives exactly the static case
Eq.~(\ref{3d-equations-physical}), i.e., the Jeans swindle
in static space corresponds {\it formally} to the non-expanding
limit of an expanding FRW universe. This static solution
$a(t)=1$ is, however, a solution to Eqs.~(\ref{hubble-law})  
and (\ref{ddot-a}) only if $\rho_0=0$ (and $C=0$), i.e., 
it is not a {\it physical} limit of the expanding case but
corresponds to the different prescription, 
Eq.~(\ref{3d-equations-physical}), for calculating
the force in the infinite volume limit. While almost all 
numerical studies are of the 
expanding case (for a review, see e.g., \cite{bagla_review}), 
a recent study \cite{sl1} of the static case for 
such initial conditions has shown that the evolution
of clustering is, in essential respects, qualitatively 
similar in both cases. This suggests that it may be possible
to understand essential qualitative features of the dynamics of
structure formation in the universe in the conceptually
simpler framework in which there is no expansion.  

With the $3-d$ equation of motion in the form of
Eq.~(\ref{3d-equations}) it is evident how the static 
$1-d$ model discussed above is naturally modified to mimic 
the $3-d$ expanding case: one can simply replace the 
force term due to the infinite $3-d$ distribution 
[i.e. the sum on the right-hand side 
of Eqs.~(\ref{3d-equations})] by that due to 
the $3-d$ distribution consisting of infinite sheets. 
The summation prescription implementing the Jeans swindle 
for the general $3-d$ case, i.e. spherical top-hat summation,
is then, as we have discussed at length above, most
appropriately replaced by the smooth prescription we have 
given. Thus we take the following $1-d$ equation for the
positions $x_i$ of the particles (sheets):
\begin{equation}
\label{sheets-3dexpansion}
\ddot{x}_i +
2H \dot{x}_i= -\frac{2 \pi G \Sigma}{a^3}
\lim_{\mu \rightarrow 0} \sum_{j\neq i}
\textrm{sgn}(x_i - x_j) e^{-\mu \vert { x}_i - {x}_j \vert}, 
\end{equation}
where the sum extends over the infinite distribution of
sheets, and we have explicitly made the identification 
$g=2 \pi G \Sigma$ (where $\Sigma$ is the mass per unit 
surface).

With initial conditions in the class of $1-d$
infinite perturbed lattices for which we have shown 
the sum for the force to be well defined and given 
by Eq.~(\ref{force-single-particle}), we then have 
\begin{equation}
\label{toy-expanding}
\ddot{u}_i +
2H \dot{u}_i= \frac{4\pi G \rho_0}{a^3} u_i \,,
\end{equation}
where we have used that the mean comoving mass 
density $\rho_0=\Sigma n_0$ (i.e.
physical mass density when $a=1$). As 
in the static case, this equation of motion remains valid at all times if
we exchange the labels of particles when they cross,
so that they bounce instead of passing through one 
another. 

For the case of an Einstein de Sitter (EdS) universe,
which corresponds to $C=0$ in Eq.~(\ref{hubble-law}),
$a(t)=(6\pi G \rho_0)^{1/3} t^{2/3}$ and 
Eqs.~(\ref{toy-expanding}) simplify to
\begin{equation}
\label{toy-expanding-EdS}
\ddot{u}_i +\frac{4}{3t} \dot{u}_i
= \frac{2}{3t^2} u_i
\end{equation}
of which the independent solutions are $u_i (t) \propto t^{2/3}$ and
$u_i(t) \propto t^{-1}$ [which are simply the well known
growing and decaying solutions for small perturbations 
to a self-gravitating fluid in an EdS universe (see, e.g., \cite{peebles})].
The evolution in between ``collisions'' is thus given by
\begin{eqnarray}
\label{expanding-evolution}
u_{i}(t) &=& u_{i}(t_0) \left[\frac{3}{5}\left(\frac{t}{t_0}\right)^{2/3}
+\frac{2}{5}\left(\frac{t}{t_0}\right)^{-1} \right] \nonumber\\
 && + v_{i}(t_0)t_0 \left[\frac{3}{5}\left(\frac{t}{t_0}\right)^{2/3}
-\frac{3}{5}\left(\frac{t}{t_0}\right)^{-1} \right]\,.
\end{eqnarray} 
Note that, from Eq.~(\ref{expanding-evolution}) 
the determination of the crossings in these 
models, instead of a quadratic equation in the static model, 
thus involves the solution of a fifth order equation (for $t^{1/3}$).

\subsection{Discussion of previous literature} 
\label{Discussion of previous literature} 

\subsubsection{Static models}

A few previous studies \cite{aurell_etal, valageasOSC_1,valageasOSC_2}
have considered static $1-d$ toy models, defining the force on the right hand 
side of Eq.~(\ref{hamiltonian}) as the derivative of a potential,
which is the sum of the contribution from the sheets in a 
finite system of size $L$, and an additional one due to a uniform  
negative background. This is exactly the ``naive'' version of
the Jeans swindle discussed above, and corresponds exactly to the 
prescription Eq.~(\ref{force-symmTH-diff}) for the
calulation of the force (with $L$ finite). The authors of 
\cite{aurell_etal} discuss explicitly the problems 
associated with taking the infinite system limit. As a result 
they limit their analysis only to a case for which their 
prescription gives a unique and finite result: a finite number of particles 
displaced off an infinite perfect lattice, modelling a finite 
localized perturbation embedded in an otherwise uniform universe.
It is simple to verify that equation of motion for these
displacements is then exactly Eq.~(\ref{inverted-oscillator}),
which we have now shown to be valid for the infinite 
lattice with perturbations which do not break the lattice translational
invariance.  

In \cite{valageasOSC_1,valageasOSC_2}, on the other hand, the 
dynamics is formulated for a system of finite $L$, and the 
problem of the definedness of the force in the infinite 
system limit is not explicitly addressed. Instead it is dealt
with implicitly by assuming that the finite system is
symmetric about some point. Taking this latter point as origin of 
coordinates, the top-hat prescription Eq.~(\ref{force-symmTH-diff}) for the force 
at coordinate position $x$ may then be rewritten as
\begin{equation}
F(x)=-2gN(0,x)+2gn_0x \,,
\label{symmetric-force}
\end{equation}
in which the size of the system does not explicitly
appear. Labelling the particles by their position with respect to
the origin ($i=1...N$), the force on the $i$-th particle
may then be written 
\begin{equation}
F_i=2gn_0\left[ x_i - \left(\frac{L}{N}\right) (i -1) \right],
\label{top-hat-symmetric-displacement}
\end{equation}
where $x_i$ is the position of the particle. For any {\it finite}
system the quantity in brackets can be considered as the 
displacement $u_i$ of the particle $i$ from its ``original'' 
lattice site [at $(i-1)L/N$]. Thus the equation of motion
for the particles is again identical to that we have
derived.  

\begin{figure}[!t]
\begin{center}
\includegraphics[width=6cm, angle=270]{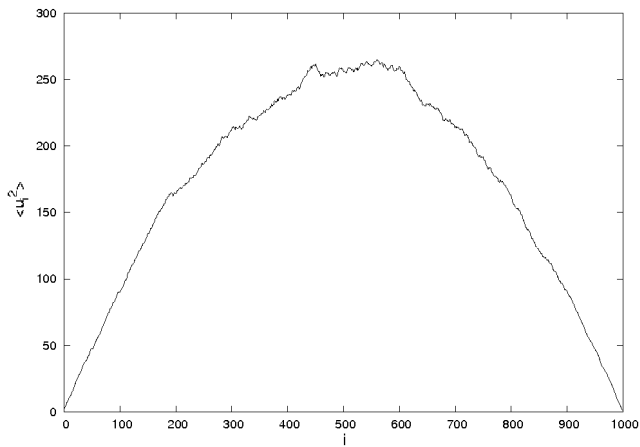}
\end{center}
\caption{The variance of the displacement $u_i$ (see text) as a 
function of a particle's ordered position $i$, calculated for one thousand
realizations of one thousand particles randomly placed in an interval.}
\label{poisson}
\end{figure}

We note again that we have derived this force law 
in this article without the assumption of symmetry (and
without the explicit introduction of a background).
Further, and most crucially, we have shown it to remain
valid for a certain class of distributions when the
infinite volume limit is taken ---  perturbed 
lattices without crossing and displacements of finite 
variance. In this respect we underline, as we have done in 
Sect.~\ref{From finite to infinite systems}, that
while in the formulation of \cite{valageasOSC_1} the same
equations of motion Eq.~(\ref{inverted-oscillator}) 
are valid for the particles in {\it any} finite
symmetric system, this does not mean that the
infinite system limit is well defined, even with
the assumed symmetry. It is illustrative to see what
``goes wrong'' when the infinite system limit is taken,
specifically, for the case of a Poisson distribution, i.e.,
when we consider a system of size $L$ in which we distribute
$N$ particles randomly, and then take
$L \rightarrow \infty$ at fixed $n_0=N/L$. The problem
is that forces, although defined at any finite $L$
by Eq.~(\ref{top-hat-symmetric-displacement}), 
diverge as $L$ does. This can most easily be seen
by considering the force written as Eq.~(\ref{symmetric-force}): the 
force on a particle at $x$, as it is proportional to the fluctuation 
in the number of particles in the interval $[0,x]$ about 
its average value, grows in proportion to $\sqrt{x}$.
Working with Eq.~(\ref{top-hat-symmetric-displacement}) this
corresponds to the fact, which can easily be shown, that 
the variance of the displacements $u_i$ (as defined above) 
diverges, violating an essential assumption for the 
perturbed lattices in the preceeding section.
Further this variance (in the finite system) depends also 
on $i$, so that discrete translational invariance, which
we also assumed, is likewise broken. These properties
are illustrated in Fig.~\ref{poisson}, which shows 
the variance of
the displacement $u_i$ as a function
of $i$, as measured in one thousand realizations of 
one thousand randomly thrown particles. In a typical realization 
the force on a particle in the centre of the box is thus much larger 
than that on a particle at the boundaries. This means that 
the typical force on a particle therefore not only
diverges as $L$ does, but that in a finite system {\it its
typical value depends on the position of the particle
with respect to the boundaries}. In practice this means 
that all the particle trajectories in a symmetric 
finite system of initially Poisson distributed particles are, 
right from the initial time, modified when $L$
is increased, and do not converge well as 
$L \rightarrow \infty$.  
Simulations of such initial conditions reported in
\cite{valageasOSC_2} show the associated coherent global
evolution of the mass distribution, which contrasts 
qualitatively with the local clustering characteristic 
of the $1-d$ (and cosmological) simulations
which we will describe in the next section. 

\subsubsection{Expanding models}

We note first that Eq.~(\ref{toy-expanding}) coincides exactly with 
that obtained in the so-called Zeldovich approximation in $3-d$ 
(see, e.g., \cite{peebles, buchert2}), when
$u_i$ is replaced by a vector function ${\bf u}({\bf x})$. This
approximation describes the evolution of displacement fields ${\bf
u}({\bf x})$ engendering small amplitude fluctuations to a
self-gravitating {\it fluid} in an expanding universe, and can be
obtained rigorously by a perturbative treatment of the full fluid
equations \cite{buchert2} in the lagrangian formalism\footnote{${\bf
x}$ is a lagrangian coordinate and the fluid is exactly uniform when
$u({\bf x})=0$.}.  For the case of one-dimensional perturbations it is
well known (see \cite{peebles} and references therein) that this
approximation becomes exact, up to the time when caustics form,
corresponding to the crossing of ``sheets'' of fluid (i.e. particles
in our case). It is thus, perhaps, not surprising, {\em a posteriori}, that
we recover exactly the Zeldovich approximation for the motion of {\it
discrete} sheets up to the time they cross: as the pair force between
sheets is independent of separation, the only way a sheet can ``see''
that the force sourcing its motion is discrete, rather than continuous
(as in the fluid limit), is when it crosses other sheets.

Eq.~(\ref{toy-expanding}) can equally be 
derived \cite{joyce_05, marcos_06} using a perturbative 
treatment of the dynamics of an infinite perturbed 
lattice (in $3-d$)  of {\it particles}. For plane wave 
displacements of the particles with a wave-vector orthogonal
to one of the lattice planes, the amplitude
of the displacement wave obeys exactly this
equation in the limit that the discreteness
of the mass distribution in these orthogonal planes 
is neglected. This latter assumption is
weaker than that used in this framework to
derive the Zeldovich approximation for a general
perturbation, which would require also that 
the displacement be of long wavelength compared 
to the discreteness scale in the direction
parallel to it.

In the studies of \cite{yano+gouda, tatekawa+maeda}, the authors study
exactly the equations of motion Eq. (\ref{toy-expanding}) for the
displacements of sheets perturbed off a perfect lattice (as in
cosmological simulations). They adopt these equations arguing that
they represent the fluid limit for $1-d$ perturbations in a $3-d$
expanding universe. While {\it before} sheet crossing (i.e. the
formation of caustics), as discussed above, this is indeed known to be
true --- these equations are just the Zeldovich approximation which
is, in this regime, exact --- the extension to longer times is argued
to be valid because the ``collisionless'' sheets of fluid will simply
pass through one another.
Our derivation of these equations shows that this in fact
corresponds to the {\it discrete} particle/sheet model.  
Indeed we have {\it not} taken the
fluid limit in our derivation, and the equations do not represent the
fluid limit of this model. It simply happens to be the case that in
this model, before crossing, the equations correspond with those in
the fluid limit, for the physical reasons we have mentioned
above. After crossing this equivalence breaks down, and the
prescription used by \cite{yano+gouda} to ``analytically continue''
the fluid model beyond its regime of validity actually maps onto this
discrete particle/shell model. We will discuss further in our
conclusions the definition of the fluid limit for the $1-d$ models we
are considering here.

The other two groups who have considered $1-d$ toy
models incorporating $3-d$ expansion have, as in this
article, worked in a particle/sheet framework.
Both the original model, proposed in \cite{rouet_etal} and 
studied further in \cite{miller+rouet_2002}, and the
subsequent one proposed and studied in
\cite{aurell+fanelli_2002a, aurell+fanelli_2002b}
and \cite{miller+rouet_2006},
derive their (different) equations of motion by
following, formally, the steps described above
leading from Eq.~(\ref{3d-equations-expanding})
to Eq.~(\ref{3d-equations}). The force on the right 
hand side of Eq.~(\ref{3d-equations-physical}) is 
simply that due to the sheets, calculated in 
the analogous manner\footnote{In \cite{rouet_etal}
the force term is simply denoted $E_i$,
without an explicit prescription for calculating 
it. It can be inferred from the description given
subsequently of the numerical simulations that 
the implicit summation is the symmetric top-hat
centred at the spatial origin. 
In \cite{aurell+fanelli_2002a, aurell+fanelli_2002b},
on the other hand, the top-hat regularisation is explicited.}, 
i.e.,
\begin{equation}
\label{1d-equations-physical}
\ddot{r}_i = 2\pi G \Sigma \lim_{L\rightarrow \infty}
\sum_{r_j \in [-L, L]} \textrm{sgn} ({r}_j - {r}_i) \,.
\end{equation}
The change to comoving coordinates, when assumed also to
rescale the mass in the sheets in the orthogonal direction
(so that $\Sigma \rightarrow \Sigma/a^2$), 
gives 
\begin{eqnarray}
&&\ddot{x}_i
+ 2H \dot{x}_i \nonumber \\
&=& \frac{2\pi G \Sigma }{a^3} \left[ \lim_{L\rightarrow \infty}
\sum_{x_j \in [-L, L] } \textrm{sgn}({x}_j - {x}_i) +
2 n_0 x_i \right]\,,\nonumber \\
\label{1d-equations-comoving}
\end{eqnarray} 
provided that $a(t)$ obeys the equation
\begin{equation}
\frac{\ddot{a}}{a}= -\frac{4\pi G}{a^3}\rho_0\,.
\label{ddot-a-1d} 
\end{equation}
As above $\rho_0=\Sigma n_0$ is the mass density (in $3-d$) 
when $a=1$. 

The Eqs.~(\ref{1d-equations-comoving}) are those adopted
by \cite{rouet_etal, miller+rouet_2002, miller+rouet_2006, 
aurell+fanelli_2002a, aurell+fanelli_2002b}. The term
which we have written on the right hand side of the
equation corresponds exactly to the prescription 
Eq.~(\ref{force-symmTH-diff}) for the calculation of 
the force. It incorporates the required subtraction
of the effect of the background, so that motion in
comoving coordinates is sourced only by perturbations
to uniformity. Just as in the static models of
\cite{aurell_etal, valageasOSC_1} discussed
above, which are obtained formally by setting $a=1$
in Eq.~(\ref{1d-equations-comoving}), this force
is well defined only if symmetry is assumed 
about the chosen origin in the point distribution.
This is indeed the assumption made in the numerical studies
of \cite{rouet_etal, miller+rouet_2002, miller+rouet_2006, 
aurell+fanelli_2002a, aurell+fanelli_2002b}. 

The difference between the models of 
\cite{rouet_etal, miller+rouet_2002} and
of \cite{aurell+fanelli_2002a, aurell+fanelli_2002b}
(studied also in \cite{miller+rouet_2006}) arises only 
in what is assumed in each case about the scale factor $a(t)$. 
The former authors impose an EdS cosmology behaviour
for the scale factor, $a(t) \propto t^{2/3}$, and 
require that it is a solution of 
Eq.~(\ref{ddot-a-1d}). While this is mathematically
consistent, it is not physically coherent: comparing
Eq.~(\ref{ddot-a-1d}) and Eq.~(\ref{hubble-law}) we
see that it corresponds to imposing a Hubble expansion
sourced by a mean density three times the physical mass
density of the sheet  (or, equivalently, assuming that 
the gravitational constant is not the same for the 
background as for the perturbations). 
Refs.~\cite{aurell+fanelli_2002a, aurell+fanelli_2002b},
on the other hand,  simply impose that $a(t)$ be the 
EdS expansion, with the right normalization. This
amounts to adding ``by hand'' a term to the derived equation
\cite{miller+rouet_2006}. While this is
not mathematically rigorous within the context of
the derivation just described, it is more appropriate
physically. Indeed it corresponds effectively to
simply replacing the Jeans swindle $3-d$ force term
in Eq.~(\ref{3d-equations}) by the prescription
Eq.~(\ref{force-symmTH-diff}). This differs  
from the ``derivation'' we have given above
for Eqs.~(\ref{sheets-3dexpansion}) only in the
form of the Jeans swindle adopted. For the case that 
symmetry about the origin is assumed, we have
the same equations of motion. 
In a finite system Eq.~(\ref{top-hat-symmetric-displacement}) 
is valid and so the equations of motion in their
numerical simulations reduce exactly to
Eqs.~(\ref{toy-expanding}).

In conclusion the equations of motion Eqs.~(\ref{toy-expanding})
are exactly the same as those used by
\cite{yano+gouda, tatekawa+maeda}, and by 
\cite{aurell+fanelli_2002a, aurell+fanelli_2002b, miller+rouet_2006}.
The only difference in practice between all these studies are
the initial conditions adopted and also the analysis of
the resultant clustering given. Rather than working in the 
cosmological time variables, the latter authors define, a new 
time coordinate $\tau= \sqrt{2/3}\ln t$. Eqs.~(\ref{toy-expanding-EdS}),
for the case of an EdS universe, then take the very simple form
\begin{equation}
\label{toy-expanding-RFcoords}
\frac{d^{2} u_i}{d\tau^{2}} +
\frac{1}{\sqrt{6}} \frac{d u_i}{d \tau}= u_i\,.
\end{equation}
In these variables the model is thus equivalent to an 
infinite set of inverted oscillators which bounce
elastically, with an additional {\it constant} damping.
Because of the fifth order equation which must be
solved to determine the crossings (now for the
parameter $t^{1/3}=e^{\tau/\sqrt{6}}$), the model 
has been dubbed the ``quintic'' model by the
authors of \cite{aurell+fanelli_2002a}.

The model of 
\cite{rouet_etal, miller+rouet_2002}, on the
other hand, is mathematically consistent but
of less apparent relevance to the ``real'' 
$3-d$ cosmological model, as the expansion 
it imposes on the sheets is not the physical 
$3-d$ expansion. Indeed in this respect
we note that, in the derivation of 
\cite{rouet_etal}, any function $a(t)$ satisfying
Eq.~(\ref{ddot-a-1d}) can be adopted with the
same consistency. The only way in fact in which 
the derivation of the $3-d$ equations can be 
rigorously adapted to $1-d$ is by using the
$1-d$ expansion law derived from 
Eq.~(\ref{1d-equations-physical}) in the limit
of uniformly distributed sheets. This is 
\begin{equation}
a(t)=1+H_0t-2\pi G n_0 t^2\,,
\end{equation}
where $H_0=H(t=0)$, i.e., free fall in a constant 
gravitational field of strength $4\pi G n_0$.
As this is very different to the $3-d$ expansion
law it is probably not a variant of the toy model
which is of practical interest.

\section{Case study:  
Infinite shuffled lattices in a static universe}
\label{Case study} 

We present in this section results of a numerical 
study of the static toy model above, starting from the simple ``shuffled
lattice'' initial conditions. Our aim here in this brief presentation
is simply to illustrate the qualitative similarity of the evolution
to that observed in the exactly analogous $3-d$ model, which
has been presented and studied in \cite{sl1} (see also \cite{sl2,sl3}).
In particular we wish to illustrate that the toy model manifests 
the ``hierarchical'' and  ``self-similar'' behaviours of the latter,
which are features also of expanding universe simulations 
in $3-d$. The studies cited above of this 
case \cite{aurell_etal,valageasOSC_1, valageasOSC_2} start from
different initial conditions and focus on evolution on much 
longer times scales in which the finiteness of the box, or
initial perturbation, is explicitly important. The study
in \cite{valageasOSC_1, valageasOSC_2} considers,
in particular, the regime of thermalisation, which is defined
only for finite systems.
On the other hand the dynamics we analyse is 
very similar qualitatively to that studied in the expanding 
models \cite{rouet_etal, yano+gouda, miller+rouet_2002,
miller+rouet_2006}. Indeed the methods of analysis we use
below are, as in \cite{yano+gouda}, the standard ones used in
cosmological simulations.
 
\subsection{Numerical simulations}

In a $3-d$ cosmological $N$-body simulation,
as we have noted, the underlying infinite physical 
system is treated numerically using the ``replica method'',
i.e., an infinite, but periodic, system is used.
The numerical integration involves calculating the
force on each of the $N$ particles by summing over 
this infinite system. Physical results should, of course,
not depend on the size of the periodic box $L$. The 
underlying reason why this is true is that the 
forces on particles converge well in the infinite
volume limit. More specifically the force on a given
particle is that due to particles in a finite region
about it. The size of this latter region is initially
of order the mean inter-particle separation, but 
increases as clustering develops in the system.
As long as the characteristic scale for this
clustering is small compared to the chosen box 
size $L$, results are, to a good approximation,
independent of $L$. A finite simulation can thus 
represent well the infinite system for a finite
amount of time.

In the $1-d$ case we have seen that for 
a class of perturbed lattices --- which are the
configurations used as initial conditions in
cosmological simulations --- the force is given
exactly as a trivial function only of the particle
displacement. Thus, to simulate numerically the evolution
of this infinite system, the step in which the force is
calculated is trivial (rather than involving the 
approximation of an infinite sum). The only question
which arises is how to treat the boundary conditions 
of the finite sub-system of this infinite system 
which one can simulate. Periodic boundary
conditions, i.e., particles which leave the finite 
interval on one side enter at the other side, 
are the evident simple choice, as
they have advantage of maintaining (discrete)
translational invariance. We could, however, easily 
use other boundary conditions (e.g. simply neglecting 
mass loss, or injecting mass in a stochastic manner
to compensate average loss), and our results
should not depend on this choice, just as 
they should not depend on the size of the
interval.

We consider here ``shuffled lattice'' initial conditions
and specifically with the PDF for the (independent) 
displacements applied to particles from their initial lattice sites:
\begin{displaymath}
 p(u) = \left\{ \begin{array}{ll}
                 \frac{1}{\Delta} & \textrm{if $u \in [-\Delta/2,+\Delta/2]$}~,\\
		 0 & \textrm{otherwise}~.
                \end{array} 
	\right.
\end{displaymath}
%
We set the initial velocities to zero. There are in this case 
thus two parameters in the model: the lattice 
spacing $\ell$ and the amplitude $\Delta$ of the ``shuffling''.
As we are treating the infinite system limit,
and the gravitational force provides itself no characteristic
length scale, there is in fact only {\it one relevant 
parameter} characterizing this class of initial 
conditions, which can be taken to be the dimensionless 
ratio $\frac{\Delta}{\ell}$ (just as in the analogous 
initial conditions in $3-d$, see \cite{sl1}).

Numerically we have simply evolved the particle positions
as given by Eqs.~(\ref{eqns-integrated-static-u})
and (\ref{eqns-integrated-static-v}) between crossings.
The subsequent crossing is determined at each time, and 
the positions and velocities of the crossing particles 
are updated accordingly. For numerical efficiency we
have implemented the optimized algorithm, using a heap, 
described in detail in \cite{noullez_etal}.

\subsection{Evolution of clustering: visual inspection}

\begin{figure*}
\resizebox{15cm}{!}{\includegraphics[width=6cm, angle=270]{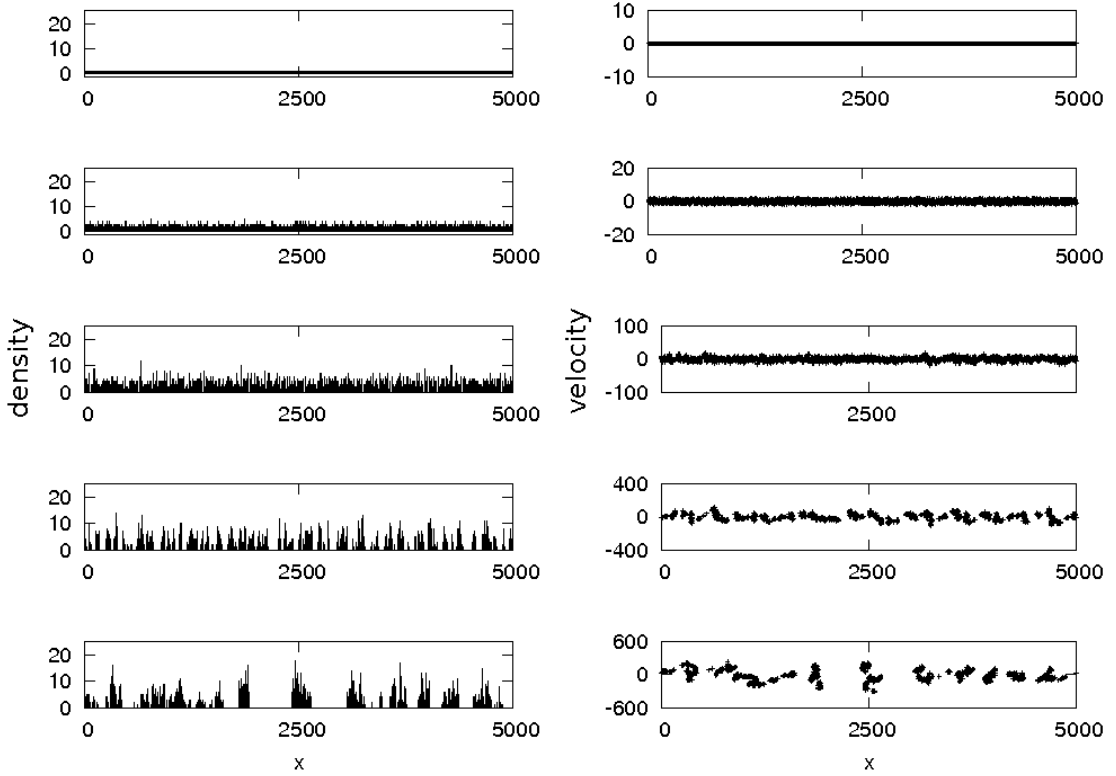}}
\caption{Left panels: the number of points in each lattice cell in the
initial conditions (first panel) and at times $t=2,5,8,10 \tau_{dyn}$;
Right panels: distribution of particles in single particle phase space
at the same times.}
\label{fig_snapshots}
\end{figure*}

Shown in Fig.~\ref{fig_snapshots} are snapshots of the initial
conditions and evolved configurations at $t=2,5,8,10 \tau_{dyn}$,
for a system with $5000$ particles. The plots
in the left hand panels show the number of particles $N(i)$ 
in each lattice cell at each time, which is proportional
to the  mass density in each cell. Defining the number density 
contrast as
\begin{equation}
 \delta(x) = \frac{n(x)-n_{0}}{n_{0}},
\end{equation}
where $n(x) = \sum_{i=1}^{N} \delta_{D}(x-x_{i})$ is 
the microscopic number density, the plots represent
the evolution of $\bar{\delta} (x) +1$, where the
bar indicates an average over the unit lattice cell.
In the phase space plots, in the right hand panels, each 
point represents simply one particle.

One sees clearly that the evolution appears to proceed
in a ``bottom-up'' manner by the hierachical formation of 
clusters which increase in size in time, starting from
the smallest scale (of order the lattice spacing). 
The sense in which the system is representative of the
evolution of an infinite system is manifest in the 
fact that the system does not appear to have a preferred 
centre --- clusters form in apparently random locations 
without sensitivity to the boundaries. Indeed we 
do not follow the evolution for longer times than those
shown precisely because the finite size of the system
then becomes important. We note that the dynamical features 
displayed in this regime are similar to those 
seen in the first studies of the $1-d$ expanding model of
\cite{rouet+feix}.


\subsection{Evolution of the Power Spectrum}

In cosmology the primary diagnostic used to characterize the evolution
of clustering in infinite space is the {\em power spectrum} (PS) (or 
{\em structure factor}) of the particle system. Since we
consider distributions which are periodic in an interval of size $L$,
we can write the density contrast as a Fourier series
\begin{equation}
 \delta(x) = \frac{1}{L}\sum_{k}\exp(ikx)\hat{\delta}(k)~,
\end{equation}
with $k \in \{(2\pi/L)n, n\in \Bbb{Z} \}$. The coefficients
$\hat{\delta}(k)$ are given by
\begin{equation}
 \hat{\delta}(k) = \int_{L}dx~\delta(x)~\exp(ikx)~.
\end{equation}
The PS is then defined as (see e.g. \cite{book})
\begin{equation}
 P(k) = \frac{1}{L} \big< \vert \hat{\delta}(k) \vert ^{2} \big>~,
\end{equation}
where $\big<..\big>$ represents the average over an ensemble
of realizations of the system.

\begin{figure}[!t]
\begin{center}
\includegraphics[width=6cm, angle=270]{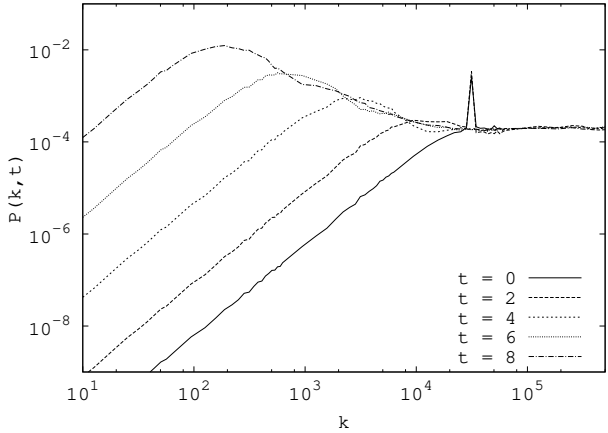}
\end{center}
\caption{Evolution of the PS, averaged over $500$ realizations of a
 system with $5000$ particles. We have set $L=1$.}
\label{PS}
\end{figure}

The evolution of the PS estimated using an average over 500 
realisations of our system (with $5000$ particles) is shown 
in Fig.~\ref{PS}.

The evolution is qualitatively like that observed in $3-d$ simulations
(both static and expanding), and the $1-d$ expanding simulations of
\cite{yano+gouda}. At small wavenumbers the evolution of the PS shows
a simple temporal amplification. This is the behaviour expected from a
linearized treatment (see, e.g., \cite{peebles}) of the equations for
a self-gravitating fluid, which gives independent evolution of each
mode of the density field governed precisely by
Eq.(\ref{toy-expanding}) in the expanding case, and
Eq.~(\ref{inverted-oscillator}) for the static case\footnote{Note
that, for the $1-d$ model the derivation of this result for the
density contrast follows trivially from the continuity
equation. The latter gives $\delta (x) =-du(x)/dx$ in the limit of small
density fluctuations $\delta (x)$, so that amplification as a
function of time of $\delta(x)$ in this limit is simply that 
of the displacements $u(x)$.}.

For the latter case, and vanishing initial velocities,
this gives
\begin{equation}
P(k,t) = P(k,0) \, \cosh^{2} (t/ \tau_{dyn} ) \,.
\end{equation}
We will see below that this is indeed a very good description of
the small $k$ evolution. Note that the initial PS at small
$k$ is a simple power law  $P(k) \propto k^{2}$, which is
that of the initial shuffled lattice. The exact form of this
PS can in fact be derived analytically (see \cite{sl1} for the 
exact expression, and \cite{book, andrea} for a derivation), and 
for the case we are considering takes the form
of a simple interpolation between this small $k$ behaviour
and the asymptotic behaviour $P(k)=1/n_0$ (characteristic
of any point distribution).

The regime in which linear amplification is valid decreases
with time, i.e., linear
amplification is observed in a range $k < k_{NL}(t)$,  
where $k_{NL}(t)$ is a wave-number which decreases as 
a function of time. 
This is precisely the qualitative behavior one would anticipate as 
linear theory is expected to hold only above a scale which, 
in real space, because of clustering, increases with time.
At all times, the PS converges at large 
wave-numbers ($k > k_{N}$, where $k_{N}$ is the Nyquist frequency) 
to the asymptotic value $1/n_{0}$. This is simply a reflection of 
the necessary presence of shot noise fluctuations at small scales 
due to the particle nature of the distribution. In the intermediate 
range of $k$, \textit{i.e.} $k_{NL}(t) < k \leq k_{N}$, 
the evolution is quite different than that given by linear theory. 
This is the regime of non-linear clustering in which the density
fluctuations are large in amplitude. 

One of the important properties of cosmological simulations is that,
starting from initial conditions with PS which are simple power law
(for $k$ smaller than the Nyquist frequency), the evolution of
clustering is, at sufficiently long times and large scales,
``self-similar'' (see \cite{sl1} for a discussion and references). 
By this it is meant that the evolution of clustering, above a given 
spatial scale, is equivalent to an appropriate
time-dependent rescaling of the length scales.
For the PS this relation is written conveniently (in
$1-d$) as
\begin{equation}
\label{self-similarity}
kP(k,t) = kR_{s}(t) \times P(kR_{s}, t_{s})\,,
\end{equation}
where $R_{s}(t)$ is the time dependent rescaling of length\footnote{ 
Note that this kind of scaling behavior of the PS is found also in other statistical
physics problems characterized by dynamical structure formation 
as for instance {\em coarsening} and {\em spinodal decomposition} in first order phase 
transitions \cite{Bray}.},
normalized by at some arbitrary time $t_{s}$. Physically this
behaviour (observed in $3-d$ for a range of initial PS) is 
interpreted as due to the fact that the non-linear clustering
at a given scale at a sufficiently long time is sensitive 
only to the initial perturbations at larger scales and
their evolution (and insensitive, notably, to the initial
interparticle distance, which provides another potential
characteristic length scale in the initial conditions).
Indeed this latter evolution (i.e. the regime of linear 
amplification) is, for a power-law initial PS, 
itself ``self-similar'' in the sense of 
Eq.~(\ref{self-similarity}), and the observation
of self-similarity is simply that this relation
extends into the non-linear regime.
The small $k$ behaviour of the PS (proportional to
$k^2$) taken together with the fact that it is amplified at 
small $k$ as given by linear theory then imply that
the self-similar scaling will be characterised
by the function
\begin{equation}
R_{s}(t) = \left( \frac{\cosh(t/\tau_{dyn})}{\cosh(t_{s}/\tau_{dyn})}
 \right)^{2/3} \rightarrow
 \exp\left(\frac{2}{3}\frac{t-t_s}{\tau_{dyn}}\right)\,.
\label{Rs-definition}
\end{equation}
It is the latter exponential form, for asymptotically large times,
which is the relevant one for the self-similar behaviour, as in
this limit the reference time $t_s$ is arbitrary.

\begin{figure}[!thb]
 \begin{center}
 \includegraphics[width=6cm, angle=270]{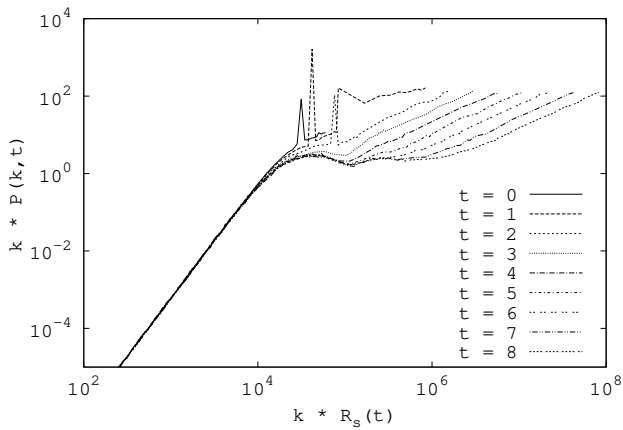}
\end{center}
\caption{Evolution of $k\times P(k,t)$ as a function of $k\times R_s(t)$ where
$R_s(t)$ is as given in Eq.~(\ref{Rs-definition}). Exact 
self-similar evolution would corresponds to the superposition 
of the curves.}
\label{fig_SSps}
\end{figure}

To assess the validity of this approximation in our system, we show in
Fig.~\ref{fig_SSps} the temporal evolution of $k\times P(k)$ as a
function of the dimensionless parameter $k \times R_s(t)$, and taking
$t_s=0$.  At small $k$, we see that right from the initial time the
self-similarity is indeed followed (as the rescaled curves are always
superimposed at these scales). This is simply a check on the result
validity of linear theory in this regime, as anticipated above.  As
time progresses we see the range of $k$ in which the curves are
superimposed increases, extending further with time into the
non-linear regime.  This is precisely what is observed in the
analogous $3-d$ simulations: as non-linearity develops it is
characterized by this self-similarity. Note that the behavior at
asymptotically large $k$ is constrained to be proportional to $k/n_0$
at all times, corresponding to the shot noise present in all particle
distributions with average density $n_0$ and which, by definition, does
not evolve in time (and therefore cannot manifest self-similarity).

\subsection{Evolution of the Mass Variance}

It is instructive also to characterise the evolution through simple
real space statistics. In order to distinguish the ``non-linear''
regime of large fluctuations from the ``linear'' regime of small
fluctuations (in which the linear fluid theory is expected to be
valid), it is useful to consider the normalized variance of particle
number (or mass) in intervals, defined as
\begin{equation}
 \sigma^{2} (x) = \frac{\big< N^{2}(x) \big> - \big< N(x) \big>^{2}}
{\big< N(x) \big>^{2}}~,
\end{equation}
where $N(x)$ is the number of particles in an interval of length $2x$.

%
\begin{figure}[!h]
 \begin{center}
 \includegraphics[width=6cm, angle=270]{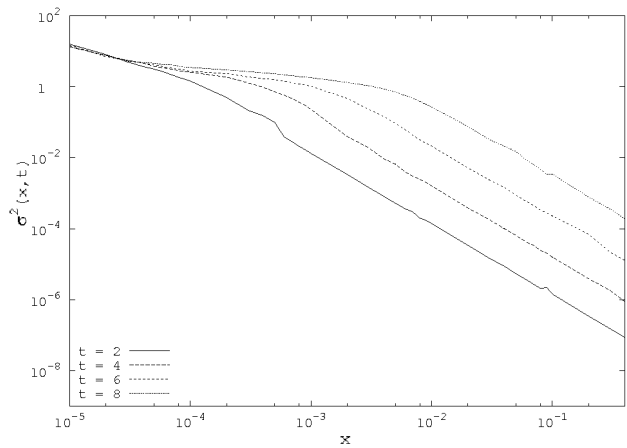}
\end{center}
\caption{Normalized mass variance in intervals of
width $2x$ for the different times indicated. As in the previous
figures the results are for an average over $500$ realizations
of a system with $5000$ particles.}
\label{variance-evol}
\end{figure}
We show in Fig.~\ref{variance-evol} the temporal evolution 
of  $\sigma^{2}(x)$ estimated using an ensemble average 
over 500 realizations of our 
system (and using periodic boundary conditions in the estimation, as 
in the simulations).
As in the case of the PS above there are three regimes. At
large scales we see a simple amplification of the initial
functional behaviour, which in this case corresponds
to $\sigma^{2}(x) \propto x^{-2}$. Note that this 
behaviour simply corresponds to unnormalized mass 
fluctuations independent of scale, which is the
most rapid decay (proportional to the
surface) possible in any spatially homogeneous point 
distribution.  At 
small scales, on the other hand, we observe
$\sigma^{2}(x) \propto x^{-1}$ which is the
shot noise behaviour intrinsic to any such
distribution at small scales. The range of 
scales between these two limiting behaviours
is that of the non-linear clustering resolved
non-trivially at this particle density. We
see that the cross-over to this regime 
from the linear regime occurs approximately
where the amplitude of the fluctuations is
or order unity.  

This behaviour of the variance illustrates very
clearly the ``hierarchical'' nature of the 
clustering, which is generic in the evolution of $3-d$
simulations starting from a very broad class of 
initial conditions: the initial small fluctuations
at a given scale are amplified (according to linear theory) 
until the fluctuations in overdense regions collapse
forming structures. Theoretically such 
behaviour is expected \cite{peebles} for any initial fluctuations
with PS with behaviours $P(k) \sim k^n$ where
$n<4$, while for  $n>4$ it is expected that the effects
of clustering at small scales will dominate over that
of the linear evolution of the very uniform
distribution at larger scales.

It is instructive to probe also in real space the self-similar 
behaviour described using the PS. To do so consider, following
\cite{sl1}, the temporal evolution of scale 
$\lambda(\alpha,t)$ defined by the relation
\begin{equation}
\sigma^{2}[\lambda(\alpha,t)] = \alpha \,,
\end{equation}
where $\alpha$ is a chosen constant. Self-similarity, at a given
amplitude of the variance, then corresponds to 
$\lambda(\alpha,t) \propto R_{s}(t)$.
\begin{figure}[!t]
 \begin{center}
 \includegraphics[width=6cm, angle=270]{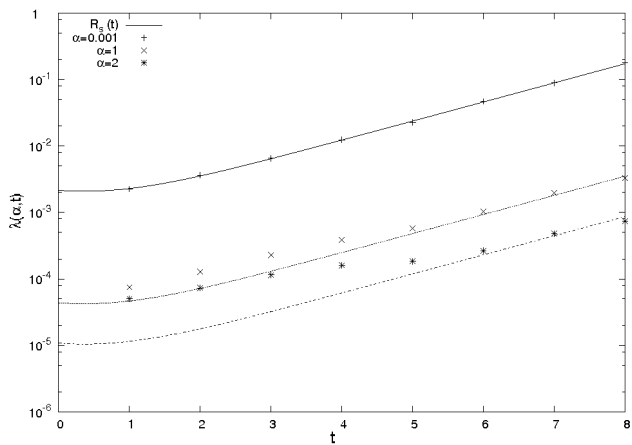}
\end{center}
\caption{Evolution, as a function of time, of the scale at which 
the normalized mass variance has the indicated amplitude. The lines
correspond to the exactly self-similar behaviour (see text).}
\label{variance_ss}
\end{figure}
In Fig.~\ref{variance_ss} we show $\lambda(\alpha,t)$ for 
the indicated values of $\alpha$, as well as the curves 
proportional to $R_{s}(t)$ corresponding to the 
self-similar behaviour. Such a graph illustrates even more
clearly than the evolution of the PS how self-similarity propagates
progessively to the regime of larger amplitudes as time 
goes on,  describing in this way the clustering further into the 
non-linear regime. 

While these behaviours are all qualitatively similar to
those in the $3-d$ case, there are also notable differences.
For example the maximal amplitude reached by the variance
in the range of times explored is much smaller than that
in $3-d$ --- of order only a few compared to a hundred
or more in the latter case\cite{sl1}. This is probably indicative
of a much weaker non-linear clustering in the former case,
associated with the smoother behaviour of the $1-d$ force
at small scales. This difference is also reflected in the
behaviour of the PS, which flattens much more rapidly
to its asymptotic Poisson value in the $1-d$ case ---
at around the Nyquist frequency rather than at
considerably larger wavenumbers in the $3-d$ case.

\section{Summary and discussion}

We have revisited in this article a basic question concerning the
definition of the gravitational force in $1-d$ infinite point
distributions. Previous definitions of this quantity in the literature
have required the assumption of the existence of a special point
(centre) in the distribution, i.e., explicit global breaking of statistical
translational invariance which is typically a feature of the infinite
distributions one instead wishes to study. We have noted that the
problem, associated with the non-converging surface fluctuations in
such distributions, may be solved by employing a definition using a
smooth screening which is sent to zero at the end of the calculation.
We have then shown explicitly that this leads to a well defined force
for a specific class of infinite perturbed lattices --- those subject
to perturbations of finite variance which do not make particles
cross. In this case, when the mean displacement of particles is also
assumed to vanish, the force on each particle take a unique value
which is simply proportional to its own displacement from its lattice
site. We note that we have assumed also that variance of the 
displacement fields is finite, which restricts to initial 
density fluctuations which have a sufficiently rapidly decaying
power spectrum at small wavenumber (specifically, such that
$P(k \rightarrow ) \sim k^{n}$ where $n>1$, analogous
to the same condition with $n > -1$  in $3-d$).

We have then discussed different dynamical toy models which
incorporate this definition of the force --- the simple conservative
Newtonian dynamics and then one which incorporates a damping term
mimicking the effect of $3-d$ expansion. Since the crossing of
particles is equivalent, up to labels, to elastic collisions with
exchange of velocities, the configurations generated by such dynamics,
at any finite time, are always in the class of infinite perturbed 
lattices for which the force is defined (provided such a configuration 
is the initial condition). This is the case because, at any finite
time, collisions/crossings may only correlate particles up to
a finite distance, and the correlation properties of displacements
at asymptotically large separations therefore always obey
the required conditions. The equations of motion are then simply those of an
infinite set of inverted harmonic oscillators (with damping in the
expanding case) with centres on the original lattice sites, and which
bounce elastically when they collide.  In this context we have also
discussed in detail the different formulations of these models in the
previous literature. We have then presented some results for the development of
clustering in the simplest model, for the simplest class of initial
conditions. We have underlined the similarity of the results to those
which have been obtained for the analogous system in $3-d$, which is
itself a simplified model for the full cosmological model. The physical 
meaning of the prescription adopted for the force is also very
clearly illustrated by these simulations: the dynamics we observe in this
infinite system limit is simply that which would occur in a system
with a screened gravitational force, in the regime in which the size
of the structures is much less than the scale of the screening.

A few additional remarks on our prescription for the
force, and the class of point distributions we have
considered, are appropriate:

\begin{itemize}

\item While we have emphasized that our definition of the force does
not require explicit breaking of translational invariance, we have
in fact shown it to give a well defined answer only for a class
of point processes which have discrete statistical translational invariance
(by lattice vectors) rather than full statistical translational symmetry.
We will consider in forthcoming work the more general question
of the general conditions on point processes for the force to 
be defined. In this respect we note that 
it has, in fact, been shown in \cite{aizenman_etal2001} that 
{\it any} point
process in $1-d$ with a variance of mass in
an interval which is bounded  necessarily breaks continuous
statistical translational invariance. While we have not
shown here that the definedness of the $1-d$ force requires, in
general, such boundedness of the variance, it is straightforward
to show that this boundedness indeed characterizes the class of  
perturbed lattices for which we have found the force to be 
defined.  

\item We note that the results of \cite{aizenman_etal2001} in fact 
generalize an earlier result (see \cite{aizenman+martin_1980}
and references therein), that 
the thermodynamic equilibria of the $1-d$ one component 
plasma (OCP, or``jellium'' model) break translational
invariance. This model is in fact just the same system 
we are considering here up to the sign of the force, with the 
difference that the presence of a physical neutralizing 
negative background is specified. We therefore do not have, 
a priori, the freedom to use the regularization of the problem in the 
infinite system limit which we have exploited here. We note, however, 
that if we were to do so, i.e., define the model in the thermodynamic 
limit as the zero screening
limit of a screened Coulomb interaction in $1-d$, we obtain,
for the class of perturbed lattices we have determined, 
a quite trivial model in which all particles are simple 
harmonic oscillators about their lattice sites, which
bounce elastically when they collide. 
It would
be interesting to investigate further whether such a 
formulation of the OCP is in fact equivalent to the usual one. 

\item Even if the force itself is not defined, with our prescription, 
for a given point process in the infinite volume limit, it may still
be possible to construct toy models of clustering in an infinite space,
with only the weaker condition that {\it differences} in forces at 
a finite separation be defined. This is related also to
the restriction on the variance of the displacements we imposed, 
which we have noted corresponds to a restriction on the small
wavenumber behaviour of the power spectra for which the force
is defined. If one considers differences in the forces rather
than forces, it is simple to show that any power spectrum
of density fluctuations which is integrable at small wavenumber
can be treated. We will discuss this issue, and the more 
general conditions for
definition of the force, in forthcoming work. 

\item In cosmological simulations initial conditions are prepared by
applying stochastic perturbations to a perfect lattice (see
e.g. \cite{couchman, discreteness1_mjbm}). These perturbations usually
have a correlated Gaussian joint PDF, which means that in the $1-d$
analogy the no-crossing condition we have required cannot be
satisfied. In practice, however, the initial amplitudes are always
chosen sufficiently small so that no such crossing occurs in the
finite simulation box, and indeed the approximation in which the
algorithm for the initial conditions is valid corresponds to this
limit. To establish rigorously that our results can be extended to
incorporate this case would require a generalisation of the
calculation we have given which incorporates the contribution to the
force PDF from crossings.  This will lead to a non-zero variance in
the PDF, but one would expect that its effect should indeed be
negligible in the limit that the typical displacement (measured by the
displacement variance) is small compared to the lattice spacing.

\end{itemize}

In future work we aim to exploit further this model,
and the expanding variant, as toy models for the 
cosmological problem. 
Despite the previous works in the 
literature which have explored various versions of
these models in the regime in which the analogy to
$3-d$ cosmological simulations is most direct,
and found various simple behaviours in the 
clustering --- notably the studies of 
\cite{rouet_etal, yano+gouda, tatekawa+maeda, miller+rouet_2002, miller+rouet_2006} --- 
they have led so far to little of the analytical insight
one might hope to gain from a toy model. For example,
although fractal behaviours have been documented
in numerical simulations by \cite{rouet_etal, miller+rouet_2002,
miller+rouet_2006}, the relevant exponents remain unexplained.
The very simple formulation of the models we have given
may help in this respect.

These toy models may also be useful in understanding other
aspects of numerical simulations in $3-d$, such as the question
of the importance of discreteness effects: in cosmological
simulations the aim is to reproduce as closely as possible
through a particle simulation the Vlasov-Poisson limit,
which corresponds to an appropriate infinite particle 
number limit. The corrections which arise due to finite
particle number are poorly understood (see \cite{discreteness3_mjbm}
for a detailed discussion and references), leading to the absence
of control on the associated errors in theoretical predictions. 
Although there are evidently important differences, the problem 
can be formulated and studied in the greatly simplified 
$1-d$ context. This requires firstly a clear formulation
of the Vlasov-Poisson limit, which so far has been given
rigorously only for finite systems \cite{braun+hepp}.
This then allows one to define an appropriate numerical
extrapolation which should be used to study convergence.
In $1-d$ there is the interesting possibility also of
performing directly simulations of the Vlasov-Poisson
system for comparison.

We thank D. Fanelli, M. Kiessling, B. Marcos, B. Miller
and F. Sylos Labini for useful discussions and exchanges, 
and B. Jancovici for providing us with 
references \cite{aizenman_etal2001, aizenman+martin_1980}. 





\end{document}